\pdfoutput=1
\documentclass[letterpaper,twocolumn,10pt]{article}
\RequirePackage[hyphens]{url}
\usepackage{usenix-2020-09}

\usepackage{tikz}
\usepackage{amsmath}
\usepackage{makecell}
\usepackage{multirow}
\usepackage{enumitem}
\usepackage{booktabs}
\usepackage{color}
\usepackage{xcolor}
\usepackage{colortbl}
\usepackage{lscape}
\usepackage[normalem]{ulem}

\usepackage[square,numbers,sort&compress]{natbib}

\usepackage[utf8]{inputenc}

\usepackage{xspace}
\usepackage{versions}

\newif\ifIsArxiv
\IsArxivtrue %

\renewcommand{\paragraph}[1]{ \medskip\noindent\textbf{#1\phantom{xxx}}}
\newcommand\hmm[1]{\ifnum\spacefactor=1001 \uppercase{#1}\else#1\fi}

\newcommand{\mean}[1]{$\mu_r$=#1}
\newcommand{\w}[1]{$\pi$=#1} %
\newcommand{\FL}[1]{FL=#1}
\newcommand{\I}[1]{I=#1}

\renewcommand{\sectionautorefname}{\S\kern-0.2em}
\renewcommand{\subsectionautorefname}{\S\kern-0.2em}
\renewcommand{\subsubsectionautorefname}{\S\kern-0.2em}

\definecolor{red1}{HTML}{D6604D}
\definecolor{red2}{HTML}{F4A582}
\definecolor{red3}{HTML}{FDDBC7}
\definecolor{neutral}{HTML}{F7F7F7}
\definecolor{blue3}{HTML}{D1E5F0}
\definecolor{blue2}{HTML}{92C5DE}
\definecolor{blue1}{HTML}{4393C3}

\newcommand{\hlq}{factor group\xspace}
\newcommand{\hlqs}{factor groups\xspace}

\newcommand{\surveyps}{survey participants\xspace}

\newcommand{\firststudyps}{\firststudy participants\xspace}
\newcommand{\firststudyp}{\firststudy participant\xspace}
\newcommand{\Firststudyps}{\Firststudy participants\xspace}

\newcommand{\Secondstudyps}{\Secondstudy participants\xspace}

\newcommand{\secondstudyps}{\secondstudy participants\xspace}

\newcommand{\interviewps}{interviewees\xspace}
\newcommand{\interviewp}{interviewee\xspace}
\newcommand{\Interviewps}{Interviewees\xspace}

\newcommand{\hunters}{hunters\xspace}
\newcommand{\hunter}{hunter\xspace}
\newcommand{\Hunters}{Hunters\xspace}

\newcommand{\hacker}[0]{hacker\xspace}

\newcommand{\bbs}{bug bounties\xspace}
\newcommand{\bb}{bug bounty\xspace}

\newcommand{\Bbs}{Bug bounties\xspace}

\newcommand{\participants}{initial study participants\xspace}

\newcommand{\firststudy}{free-listing study\xspace}
\newcommand{\Firststudy}{Free-listing study\xspace}

\newcommand{\secondstudy}{factor-rating study\xspace}
\newcommand{\Secondstudy}{Factor-rating study\xspace}

\newcommand{\bbprogram}{bug-bounty program\xspace}
\newcommand{\bbprograms}{bug-bounty programs\xspace}
\newcommand{\Bbprogram}{Bug-bounty program\xspace}
\newcommand{\Bbprograms}{Bug-bounty programs\xspace}

\newcommand{\bbplatform}{bug-bounty platform\xspace}
\newcommand{\bbplatforms}{bug-bounty platforms\xspace}

\newcommand{\Bbplatforms}{Bug-bounty platforms\xspace}

\newcommand{\manager}{\bbprogram manager\xspace}
\newcommand{\managers}{\bbprogram managers\xspace}

\newcommand{\lowskill}{beginner\xspace}

\newcommand{\cm}[1]{}

\usepackage[size=1.525cm,textsize=tiny, colorinlistoftodos, disable]{todonotes}

\newcommand{\change}[2]{\cm{\textcolor{red}{\sout{#1}}}#2}

\newcommand{\ie}[0]{{i.e., }}
\newcommand{\eg}[0]{{e.g., }}

\ifIsArxiv
    \usepackage{fancyhdr}
\else
\fi

\begin{document}

\ifIsArxiv
\thispagestyle{plain}
\fi

\date{}

\title{Bug Hunters' Perspectives on the \\Challenges and Benefits of the Bug Bounty Ecosystem}

\author{
{\rm Omer Akgul}\textsuperscript{$\diamond$}\\
akgul@umd.edu \\
\and
{\rm Taha Eghtesad}\textsuperscript{$\ast$}\\
teghtesad@psu.edu
\and
{\rm Amit Elazari}\textsuperscript{$\mathsection$}\\
aelazari@berkeley.edu
\and
{\rm Omprakash Gnawali}\textsuperscript{$+$}\\
gnawali@cs.uh.edu
\and
{\rm Jens Grossklags}\textsuperscript{$\mathparagraph$}\\
jens.grossklags@in.tum.de
\and
{\rm Michelle L. Mazurek}\textsuperscript{$\diamond$}\\
mmazurek@umd.edu
\and
{\rm Daniel Votipka}\textsuperscript{$\ddagger$}\\
dvotipka@cs.tufts.edu
\and
{\rm Aron Laszka}\textsuperscript{$\ast$}\\
laszka@psu.edu
\and
\textsuperscript{$\diamond$}\textit{University of Maryland}\hspace{1cm}\textsuperscript{$\ast$}\textit{Pennsylvania State University}\hspace{1cm}
\textsuperscript{$\mathparagraph$}\textit{Technical University of Munich}\hspace{1cm}\\ \textsuperscript{$\mathsection$}\textit{University of California, Berkeley}\hspace{1cm}
\textsuperscript{$+$}\textit{University of Houston}\hspace{1cm}
\textsuperscript{$\ddagger$}\textit{Tufts University}
}

\maketitle

\begin{abstract}
Although researchers have characterized the bug-bounty ecosystem from the point of view of platforms and programs, minimal effort has been made to understand the perspectives of the main workers: bug hunters. To improve bug bounties, it is important to understand hunters' motivating factors, challenges, and overall benefits. We address this research gap with three studies: identifying key factors through a free listing survey ($n$=56), rating each factor's importance with a larger-scale factor-\change{ranking}{rating} survey ($n$=159), and conducting \change{}{semi-structured} interviews to uncover details ($n$=24). Of 54 factors that bug hunters listed, we find that rewards and learning opportunities are the most important benefits. Further, we find scope to be the top differentiator between programs. \change{Unfortunately, we also find many challenges.}{Surprisingly, we find earning reputation to be one of the least important motivators for \hunters. Of the challenges we identify,} communication problems, such as unresponsiveness and disputes, are the most substantial\change{challenges}{}. We present recommendations to make the bug-bounty ecosystem accommodating to more bug hunters and ultimately increase \mbox{participation} \change{}{in an underutilized market}.
\end{abstract}

\section{Introduction}
\label{sec:intro}
Traditionally, organizations relied on internal security experts (e.g., red teams) and outsourced experts (e.g., penetration testing) to discover vulnerabilities in their products.
In contrast, \bbprograms---also known as vulnerability-reward programs or ``crowd-sourced'' security---incentivize independent security experts to evaluate the security of an organization's products and report vulnerabilities in exchange for rewards (financial or otherwise, such as \change{reputation}{the learning opportunity}). \change{}{\Bbprograms{} were initially spearheaded} by Netscape in 1995~\cite{ellis2022bounty}; now, many \change{large technology }{}companies (e.g., Google, Apple) and governmental agencies (e.g., U.S. Department of Defense) run \bbprograms.

However, due to their crowd-sourced nature, \bbprograms also suffer from inefficiencies.
\Bbprogram{}s may receive invalid or duplicate reports, wasting effort~\cite{Zhao2017devising,atefi2023benefits}. Further, \change{there exists competition between programs}{programs compete} to attract productive \hunters, and older programs struggle to maintain a \hunter pool~\cite{maillart2017given, sridhar2021hacking}.
On the other hand, bug hunters \change{}{(hereafter, \emph{\hunters})} face \change{similar }{}uncertainties regarding their findings and rewards, and are often disappointed by program responses~\cite{albergotti2021apple, salter2021apple}.
To mitigate some of these issues, \bbplatforms such as  Bugcrowd~\cite{Bugcrowd} and HackerOne~\cite{HackerOne} have emerged, connecting \bbprograms to \hunters through a marketplace. However, \change{even on platforms, }{}many issues persist.

\change{}{Perhaps due to these issues, despite many benefits, bug-bounty adoption by organizations has remained relatively low\footnote{\change{}{The two largest \bbplatforms host $\sim$3,300 (global) \bbprograms~\cite{bugcrowd2020report,hackerone2020report} compared to more than 10,000 ``software publishers'' and 60,000 ``custom computer programming services'' in the U.S. alone~\cite{census2019companies}. Most (80\%) of the Forbes 500 have no vulnerability disclosure program~\cite{tod2020icer}.}} and well below predictions (as cited in \cite{bugcrowd2018growth, sridhar2021hacking}). Further, only a small fraction of \hunters find a substantial amount of bugs \cite{ellis2018empbounty, luna2019productivity, walshe2020empirical} despite the seemingly populous talent-pool reported~\cite{hackerone2020report,bugcrowd2020report}. These \hunters tend to receive more attention from the ecosystem, while others are ignored~\cite{ellis2022bounty}.}

\change{}{Identifying specific factors that make \bbprograms (un)attractive and (un)successful, addressing the most significant challenges, and bolstering commonly enjoyed benefits in the bug-bounty ecosystem could invite more \hunters, increase their commitment, enable identification of more bugs, reduce wasted effort in bug hunting and reporting, and streamline the process of fixing reported bugs; all of which could improve the security posture of many companies and improve software security more broadly. Additionally, by understanding \hunter motivations, we can also understand the societal impacts of the \bb market and guide the efforts of regulators to ensure the market meets society's broader needs.}

\change{}{To improve the \bb ecosystem, we must first understand how \bbs work.} Indeed, a number of research efforts have taken steps in this direction~\cite{finifter2013empirical, zhao2015empirical, maillart2017given, laszka2018rules, luna2019productivity, elazari2019private, walshe2020empirical, sridhar2021hacking, laszka2016banishing}.
However, a common limitation is that researchers consider data collected only from the perspective of \bbprograms (e.g., vulnerability reports and payments).
Therefore, they provide only a limited view of bug hunters' work, \change{merely considering their output and}{considering only final outputs but} neglecting \change{}{the \hunters'} motivations and the challenges that they face.

Recent work has begun to consider decision-making in vulnerability discovery broadly~\cite{votipka2018hackers,fulton2023v4a,VotipkaREObservations2020}; but none of this work focuses on \bbs specifically, discussing them only when broached by participants. Additionally, none of this hacker-focused work has empirically evaluated factors that govern \hunters' choices of which software to evaluate. %

\change{}{\Bbplatforms have themselves issued several reports on \hunters' motivations~\cite{bugcrowd2021report, bugcrowd2020report, bugcrowd2018report, hackerone2019report,hackerone2020report,hackerone2021report}.} However, 
these \change{}{brief} reports do not \change{consider}{focus on} challenges faced by \hunters, appear to be for marketing, and are not independently verified. 

\change{Though few researchers have shared similar research goals to ours~\cite{alomar2020nicebugs, ellis2022bounty}, their research remains qualitative and broad.}{Other research efforts have used interviews with stakeholders throughout vulnerability disclosure, including bug bounty programs, to explore drawbacks of the gig-work model~\cite{ellis2022bounty} and challenges in the vulnerability discovery ecosystem as a whole~\cite{alomar2020nicebugs}. We build upon these broad, qualitative studies with a larger, mixed-methods sample that explicitly and systematically identifies and quantifies the factors that affect \hunters' participation in \bbprograms. Unlike prior work, we ask \hunters to quantify how important individual factors are, allowing stakeholders to know which factors to prioritize. For instance, like other researchers~\cite{ellis2022bounty,walshe2020empirical}, we find reputation to be a motivator; however, we are able to demonstrate that it is in fact one of the least important motivators (\autoref{sec:rewards}).}

\change{In this paper, we take a step toward understanding \hunter motivations and challenges with bug bounties. Specifically, we set out to answer the following research questions}{Our specific research questions are as follows}:
\begin{enumerate}[label={\bf RQ\arabic*:},itemsep=-0.5em,topsep=0em,leftmargin=2.8em]
	\item What are the factors\change{and issues}{} that \hunters consider \change{}{and challenges they face} when participating in \change{the \bb ecosystem}{\bbs}?
	\item How important are these factors to \hunters? Why?
     \change{ \textbf{RQ3:} Do \hunters cluster based on preferred factors? }{}
\end{enumerate}

\change{With the growth of \bbs, \hunters{} have a variety of programs to choose from. With our first two questions, we seek to enumerate and stratify the broad range of factors influencing participation in \bbs.}{ 
We approach these two questions in four contexts: (1) factors considered when choosing between specific programs, (2) challenges faced, (3) benefits of \bbs in general, and (4) useful features of \bbplatforms.}

\change{Expecting a wide variety of factors considered by \hunters, in our final research question we ask if there are different archetypes of \hunters. We were particularly curious if skill level influences factor priorities.}{}
We conduct three studies (\autoref{sec:method}) to address our research questions: an initial factor-identification survey ($n$=56) to list prominent factors at play (RQ1), a larger survey ($n$=159) to find the importance of the factors (RQ2) 
and a \change{n}{semi-structured} interview study to reveal why factors are important ($n$=24).

As expected, we find the most salient benefits to be monetary\change{}{, both when choosing between programs and as a general motivator} (\autoref{sec:rewards}).
\change{}{When choosing between programs,} \hunters consider heuristics (e.g., scope) that increase the probability of finding a bug (\autoref{sec:finding_bugs}). Aside from monetary benefits, \hunters deeply value learning opportunities \change{, including learning by doing, interacting with the community, and studying examples presented through public disclosures of bugs} (\autoref{subsec:learning}). \change{}{Contrary to intuition from prior work~\cite{sridhar2021hacking, ellis2018empbounty, walshe2020empirical}, we find that they value reputation much less than other factors.}

We observe that \hunters{}' most prominent challenges are communication issues
with \managers who grade reports and decide on payouts. Specific issues include poor responsiveness, bug-grading disputes, and \change{inadequate mediation}{dissatisfaction with mediation and platform triaging} (\autoref{subsec:disputes}).

\change{Finally,}{} We also find that the gig-work model of bug bounties introduces unique challenges for \hunters. While it provides flexibility,
it can also create stress and uncertainty (\autoref{sec:freelance}).

\change{}{We discuss \bbplatform features that \hunters consider most useful: public dashboards and easy procedures for reporting bugs and receiving payments (\autoref{subsec:platform}).}

\change{}{Next, our results show the importance of legal safe harbors to \hunters in multiple contexts (\autoref{sec:legal_safe_harbor}).}
\change{While we observed general relationships among the factors identified, there was little consensus among \hunters{} on the reasons for any factor's significance. }{}
\change{However, our results hint at differences based on skill levels, which should be tested in future work (\autoref{sec:skill_levels}). 
Specifically, we observed that lower-skilled \hunters prioritize advancement opportunities and educational value in their decision-making. Higher-skilled \hunters emphasize \bbprograms that pay well for high-severity bugs, likely because they expect they are more likely to find these complex bugs, and program responsiveness, likely because they expect to submit a large number of bug reports.}{}

\change{Finally, we list suggestions \bbprograms{} and platforms can focus on to support and attract \hunters (\autoref{sec:discussion}).}{The paper concludes with a discussion of how our results expand our understanding of benefits and challenges in \bbs, as well as recommendations for a \bb ecosystem that works better for bug hunters, companies, and software security in general (\autoref{sec:discussion}).}

\section{Related Work}
\label{sec:related}
Researchers have tried to understand \hunter{}s' motivations through empirical analysis of market behaviors and via direct surveys and interviews with \hunters.

\paragraph{Market behaviors} To understand how \hunters{} select \bbprograms{}, researchers have studied empirical data produced by \bbprograms{} (e.g., vulnerability reports and payments)~\cite{finifter2013empirical,zhao2016crowdsourced,huang2016,magazinius2019we,ruohonen2018bug,alexopoulos2021vulnerability,sridhar2021hacking,walshe2020empirical,maillart2017given}. These studies investigate the relationship between \hunter{} activity and various program features, highlighting correlations that might suggest motivations. For example, researchers found that \hunter{} program selections were associated with expected monetary rewards and program age~\cite{laszka2018rules, maillart2017given}. These results are in line with similar investigations of public reporting from the Google Chrome and Mozilla Firefox \bbprograms{}~\cite{finifter2013empirical} and public HackerOne data~\cite{maillart2017given}. Though we \change{can confirm these results}{report some overlapping factors}, our work \change{differs}{offers a substantially different perspective,} as we survey \hunters directly, allowing them to tell us their priorities directly rather than inferring them.

\paragraph{\Hunters{}' self-reported motivation} Other \change{work}{publications} have leveraged surveys to characterize \hunter{} demographics and motivations. The most prominent examples of this work are \change{}{marketing materials produced annually} by HackerOne~\cite{hackerone2020report,hackerone2021report} and Bugcrowd~\cite{bugcrowd2018report,bugcrowd2020report, bugcrowd2021report}, the two largest \bbplatforms{}. Each company surveys the \hunters{} participating on their platform, collecting demographics and a high-level view of \bb{} participants' motivations (e.g., money, education). However, these surveys do not provide the same depth of exploration into \hunter{} motivation as our work and do not focus on challenges faced by \hunters{}. 

Perhaps most related to our work is a \change{}{non-profit research organization's} interview study of \change{40 \hunters{}}{\bb ecosystem stakeholders} to understand their experiences~\cite{ellis2022bounty}. \change{Similarly to our work, they identified}{The authors suggest but do not systematically define} several benefits and challenges for \hunters{}, with a \change{focus primarily on the challenges of}{primary focus on criticizing} the gig-work model. \change{}{Our work differentiates itself by employing mixed methods to systematically identify, define, and \textit{quantify} factors relevant to \bbs under four contexts: choosing between \bbprograms, challenges of \bbs, benefits of \bbs, and useful features of \bbplatforms.} Our quantification of relevant factors enables stakeholders of the \bb ecosystem---including \bbplatforms but also \hunters, companies seeking to improve their security, and potentially regulators or standards bodies concerned with security---to make better informed decisions. 

A more general study \change{investigating}{explored} the vulnerability disclosure process as a whole, finding communication to often be an issue~\cite{alomar2020nicebugs}. \change{}{Fulton et al.\ explored issues marginalized groups face in the vulnerability discovery space (e.g., women, people of color)~\cite{fulton2023v4a}. While both studies do touch briefly on hunters' perceptions of \bbs, it is tangential to their research.}%

\section{Method}
\label{sec:method}
We designed and conducted three studies to investigate our research questions: an initial \textit{\firststudy} to determine factors at play (RQ1), a \textit{\secondstudy} (RQ2), and finally an \textit{interview study} (RQ2). The first two studies allow us to understand what motivates and challenges \hunters, while the interview study contextualizes these results.

Our institutions' ethics review boards approved all three studies. Participants signed consent forms detailing study plans and participant rights before data collection. Identifiable data was only available to authors named on the ethics review.

\subsection{\Firststudy (RQ1)}
\label{subsec:firststudymethods}
To identify factors that influence participation in \bbs, we performed an online survey on \hunters ($n$=56).

\paragraph{Survey}
The survey began with open-ended questions asking participants to list factors that affect them in five \change{}{(later reduced to four, see end of~\autoref{subsec:firststudymethods})} contexts (we call these \textit{\hlqs}): (1) factors when choosing between \bbprograms, (2) reasons for leaving \bbprograms, (3) the benefits of participating in \bbs, (4) \change{the challenges they face}{challenges faced} in general, and (5) useful features of \bbplatforms. 
 
For each question, we stressed that \participants should list all factors they may consider, even if they do not regard a given factor in every single decision. Additionally, we asked \participants to spend time to recall factors if they thought there might be more they could remember. Common in listing exercises, this prompt allows us to elicit less obvious factors~\cite{bernard2017research}. We use an open-ended listing approach, called \textit{free listing}, common in anthropological research when the domain is not well understood~\cite{bernard2017research}. This is useful for eliciting the full breadth of possible factors.

Next, we asked participants to self-report their bug-bounty experience (\change{e.g., estimated number of bugs discovered, revenue earned, and years of experience}{see~\autoref{tab:demo}}) and skills. Finally, we concluded with standard demographic questions (\change{e.g., educational background, age group, country of residence}{see~\autoref{tab:demo}}) to understand our sample population. We also asked if participants were willing to be contacted for follow-up studies. The full survey can be found in \ifIsArxiv \autoref{app:firststudysurvey}\else the extended paper~\cite{akgul2023bug}\fi.

\paragraph{Pilots}
Through personal connections, we recruited three security experts who regularly work on \bbs. We sent them the survey and discussed responses in an online focus-group session. We proceeded with data collection once the questions were clear and provided good face validity~\cite{gravetter2018research}.

\paragraph{Recruitment}
We recruited by advertising on social media (through the authors' accounts), mailing lists, and Slack channels that \hunters use. In total, we received 61 complete responses to the survey. We removed 5 responses due to poor quality (unintelligible answers, unreasonably fast completion times, or duplicates), leaving 56 responses for analysis. Responses were obtained from May to December~2019.

We concluded data collection after the final 15 responses largely confirmed the factors identified in the first 41 responses, indicating conceptual saturation~\cite{charmaz2006constructing}.

\paragraph{Data Analysis}
\label{sec:analysis}
We analyzed open-ended survey responses with exploratory open coding~\cite{saldana2021coding}. Because we planned to use the identified factors directly in the \secondstudy, we calculated inter-rater reliability~\cite{mcdonald2019reliability}.

\change{In some cases,}{Sometimes} factors listed were \change{unclear}{polysemous}. In cases where \change{unclear}{these} responses came 
from participants who agreed to an interview, we asked for clarification \change{during the interview} ($n$=7). %

We developed the codebook and established reliability on the first 41 responses (73.2\% of all responses, 64.0--78.2\% of all listed factors per question\footnote{Responses are \textit{unitized} based on number of factors listed in a response (i.e., codes are assigned to individual factors, not entire responses). 
}).
The initial codebook was developed by three researchers using 10 responses (25\% of the responses at the time; 15.3--28.0\% of factors listed). Two of the three researchers then attempted to establish good reliability by independently coding batches of 10 responses at a time, resolving differences and updating the codebook after each batch. We ran out of new responses without being able to establish our threshold for acceptable reliability (Cohen's $K > 0.8$). However, after 28 days, the researchers revised the codebook and independently re-coded 16 of 41 responses ($\sim$40\% of the responses at the time; 27.0--38.3\% of all factors listed), achieving ``almost perfect''~\cite{landis1977measurement} reliability (Cohen's $K > 0.8$ for all \hlqs; 0.81--0.91). Finally, with reliability established, one researcher re-coded the rest of the responses. The final 15 responses were received after reliability had been established and coded by one researcher. The final codebook contained 78 factors across five \mbox{\hlqs}.

\paragraph{Refining the factors} 
Our listing exercise identified many factors, so it was impractical to ask participants to respond to each directly in the \secondstudy. In addition, we noticed significant overlap between two \hlqs: reasons for quitting a \bbprogram, and challenges faced in the \bb context more generally. To reduce the number of questions asked, we merged the two, resulting in 54 factors across four \hlqs (see~\autoref{tab:refined_codes}).

\subsection{\Secondstudy (RQ2)}
\label{subsec:ratingmethods}
The \firststudy identified a comprehensive set of factors considered by \hunters, but not their relative importance. %
We therefore designed a second study asking participants to rate how important each factor is to them, personally.

\paragraph{Survey}  
The second survey started with \surveyps rating each of 
the 54 factors on a seven-point Likert.\footnote{``Extremely challenging/important'' to ``Not at all challenging/important'' or ``Extremely useful'' to ``Extremely useless'' as appropriate.} As in the prior survey, the factors were asked in the context of their respective \hlqs: factors considered when choosing a program, challenges faced and benefits of \bbs, and useful bug-bounty-platform features. 
To better understand \hunters' background, we asked an open-ended question on how they started \bbs{}. As before, we finished with questions about the participant's bug-hunting experience and demographics. We again asked whether participants would be willing to 
be contacted for follow-up studies. 
The full survey can be found in \ifIsArxiv \autoref{app:secondstudysurvey}\else the extended paper~\cite{akgul2023bug}\fi.

\paragraph{Piloting}
We piloted the survey on five \hunters, focusing on how well participants understood the provided list of revised factors. We made minor revisions to the naming and explanation of the 54 factors based on these comments.
The final list of factors to be evaluated appears in~\autoref{tab:refined_codes}.

\paragraph{Recruitment} In addition to the methods used in the \firststudy, we reached out to
586 \hunters who had a public bug report in the first half of 2020 on HackerOne or BugCrowd, listed their Twitter accounts, and allowed direct messages from anyone. The vast majority of responses likely came through this method. Further, we advertised on \texttt{reddit.com/r/bugbounty}. Of 161 completed survey responses, we discarded two with nonsensical responses to multiple open-ended questions, leaving 159 responses for analysis. \change{}{We estimate there to be 4-6 overlapping participants between the free-listing and \secondstudy.}

\paragraph{\change{}{Data analysis}}
\change{}{A straightforward (and common) 
way of analyzing which factors are the most important would be to convert our participants' Likert choices to numeric values and present simple averages. This approach, while seemingly intuitive, has been criticized and discouraged by statisticians~\cite{dittrich2007paired,lubke2004applying,wu2007empirical}, because it makes the dubious assumption that the ordinal options that participants select are equidistant (e.g., the distance between ``Extremely important'' and ``Very important'' is the same as between ``Very important'' and ``Moderately important''). Thus, we adopt comparison-based techniques, which consider only whether one factor is rated higher than another. Specifically, we employ log-linear Bradley-Terry (LLBT) modeling to synthesize \emph{worth} estimates ($\pi$) that represent the relative importance of factors ($f$) to participants on a preference scale~\cite{dittrich1998modelling, dittrich2007paired}. The probability of one factor being preferred over another is given by:} 
\setlength{\abovedisplayskip}{4pt}
\[\change{}{p(f_j>f_k|\pi_j,\pi_j)=\frac{\pi_j}{\pi_j+\pi_k}}\]
\setlength{\belowdisplayskip}{0pt}
\change{}{where $j,k$ denote the indices of factors considered~\cite{dittrich2009fitting}.\footnote{We use a version of LLBT 
that requires converting our data to explicit paired comparisons. Methods without this conversion are computationally infeasible for our high-dimensional data~\cite{dittrich2007paired, hatzinger2012prefmod}. Researchers who developed these statistical methods confirmed that our approach is appropriate~\cite{akgul2022privatecomm}.
}}

\subsection{Interviews}
\label{subsec:interviewmethods}

We invited all \change{}{consenting} \firststudyps with valid responses to take part in remote semi-structured interviews. We asked if and why the identified factors were important to them. The interviews ($n$=8) started by asking how participants got into \bbs and security in general. Next, we explained each \hlq to provide context and then asked the participant to pick the most important factors and explain their choices. Finally, we asked whether and how they would continue participating in \bbs.

Similarly, we invited \change{participants in the \secondstudy}{\secondstudyps} who consented to interviews, with minor revisions to the interview protocol. Specifically, we used the revised list of factors (54 items), and asked participants ($n$=16) directly about the factors they rated highly in their survey responses. 
\change{}{In both interview rounds, interviews started while survey recruitment was still active in order to retain a higher percentage of participants.}

It was infeasible for interviews to cover all \change{possible}{54} factors. Thus, we focused on factors the participant cared about the most, \change{adding questions}{asking} about \change{less important}{other} factors \change{only}{}if time 
allowed. Interviewers were careful to avoid asking redundant questions, as \change{questions about one factor frequently spurred discussion about other factors as well}{discussion on one factor frequently expanded to others.}

With 54 factors to cover and many \hunters expressing unique considerations, reaching saturation on participants' opinions of each factor was not realistic; instead, we aimed to collect enough data to contextualize the survey results.

We conducted interviews with 24 people in total, each \change{lasting}{averaging}  43 minutes\change{on average}{}. While survey participants were not compensated,\footnote{\change{}{Our varied recruitment methods, international participants, and relatively short surveys made compensation logistically difficult.}} \interviewps were thanked with a \$20 gift card.

\paragraph{Analysis}
We again used exploratory open coding to analyze our interviews~\cite{saldana2021coding}. Two researchers coded three interviews to create an initial codebook. They then independently coded three interviews at a time, meeting to discuss the interviews, resolving differences, and updating the codebook. The final codebook was obtained when all interviews were coded and discussed. Interviews were re-coded with the final codebook,
resulting in a total of 1004 coded segments in the 24 interviews. We did not seek inter-rater reliability metric, since we see the interviews primarily as adding context to the outputs of the \firststudy and \secondstudy ~\cite{mcdonald2019reliability}.

\subsection{Limitations}
\label{sec:limitations}

Our methods are primarily based on self-report data, which is subject to well-known limitations. Self-report data often has high levels of noise and therefore does not ensure definitive answers through one measurement. We therefore investigated our overall research goal through three distinct studies, and noted the few inconsistencies that occurred. 

Lack of recall can be an issue in free listing~\cite{annett2003hierarchical}. We prompted participants to try to recall all possible factors and only switch to the next question when they could no longer think of any, a best practice for free-listing recall~\cite{bernard2017research}.

\Hunters might have portrayed themselves as more successful than they are. \change{}{Similarly, our gift card incentive might have been most attractive to interview participants who make less money in \bbprograms, perhaps due to lower skill or experience.} We partially addressed this by \change{}{recruiting the majority of \firststudyps from non-public \hunter messaging channels and \secondstudyps from \hunters from authors of publicly disclosed bugs, implying some level of expertise. Further, the metrics we collected indicate a relatively even distribution across all skill and experience levels (see~\autoref{tab:demo})}.

\change{}{On the other hand, it is likely we did not capture people who wanted to participate in \bbs but were unable to at all. This problem of survivor bias is likely unavoidable, as there is no clear way to recruit people interested, but not active in \bbs. However, we expect some of our participants with less \bb participation will have similar experiences.}

Further, all three of our studies, like all self-report studies~\cite{breen2022large}, include some sampling and selection biases. \change{We likely limited bias concerns by recruiting participants from a diverse set of backgrounds (various educational attainment, expertise, countries etc.).}{Our sample is likely reasonably representative of the current \hacker population (\autoref{sec:demographics}) and includes diversity in participant location, education, experience, and skill, meaning our results are reasonably likely to generalize to the average \hunter. However, the \hunter population itself is demographically skewed (e.g., young, white or Southeast Asian, and male~\cite{hackerone2020report,bugcrowd2020report, bugcrowd2021report}), likely introducing inherent biases 
(e.g., the companies and vulnerabilities that get attention). 
For a discussion of marginalized populations, we refer readers to the work of Fulton et al.~\cite{fulton2023v4a}.}

Finally, to maximize face validity, we rigorously piloted each stage of the study\change{}{, revising procedures with feedback}.

\section{Participants}
\label{sec:demographics}

\begin{table}[ht!]
\centering
\footnotesize
\resizebox{\linewidth}{!}{
\begin{tabular}{l  l r r r}
\toprule
\midrule
& & FL & FR & I \\
\midrule
\textbf{Gender}         & Male                    &   51  & 147 & 24\\
                        & Female                  &   2  & 2 & 0\\
                        & Self-described          &   1  & 2 & 0\\
\midrule
\textbf{Age}            & 18-29                   &    34  & 124 & 16\\
                        & 30-39                   &    16  & 27 & 7\\
                        & 40-49                   &    4 & 3 & 0\\
                        & 50-59                   &    2 & 0 & 0\\

\midrule
\textbf{Residence}      & North America      & 21  & 22 & 3 \\
                        & South Asia         & 14 & 64 & 7\\
                        & Europe             & 14 & 23 & 9 \\
                        & Southeast Asia     & 2 & 14 & 0\\
                        & Middle East        & 0 & 18 & 0\\
                        & Other              & 5 & 11 & 4\\

\midrule
\textbf{Education}
                                & $\leq$ Completed H.S.         & 18  & 53 & 11 \\
                                & Trade/technical/vocational      & 1 & 3 & 0\\
                                & College, no degree 		  & 13 & 17 & 2\\
                                & Associate's degree              & 3  & 5 & 1\\
                                & Bachelor's degree               & 15 & 56 & 3\\
                                & Professional/MS/PhD                 & 5 & 19 & 4\\
                                
\midrule
\multirow{2}{.35\linewidth}{\change{}{\textbf{Weekly Hours Working on Bug Bounties}}}

                                & <5 & 13 & 34 & 3 \\
                                & 5-10 & 18 & 53 & 8 \\
                                & 10-20 & 12 & 40 & 8 \\
                                & 20-30 & 8 & 18 & 2 \\
                                & 30-40 & 1 & 13 & 2 \\
                                & >40   & 4 & 3 & 0 \\

\midrule
\multirow{2}{.35\linewidth}{\change{}{\textbf{Total Number of Bugs Found}}}
                                & <10 & 6 & 39 & 2 \\
                                & 10-99 & 18 & 65 & 10 \\
                                & >= 100 & 12 & 52 & 8 \\

\midrule
\multirow{2}{.35\linewidth}{\change{}{\textbf{Years Working on Bug Bounties}}}

                                & <1 & 2 & 3 & 1 \\
                                & 1-2 & 18 & 87 & 12 \\
                                & 3-5 & 15 & 46 & 4 \\
                                & 6-9 & 6 & 14 & 4 \\
                                & >= 10 & 0 & 5 & 0 \\

\midrule
\change{}{\textbf{Skill}}
                                & 1 - Fundamental & 3 & 9 & 1 \\
                                & 2 - Novice & 8 & 23 & 3 \\
                                & 3 - Intermediate & 23 & 68 & 9 \\
                                & 4 - Advanced & 10 & 48 & 6 \\
                                & 5 - Expert & 12 & 13 & 4 \\
\midrule
\multirow{2}{.35\linewidth}{\textbf{\change{}{Yearly Income From Bug Bounties}}}
                                & <\$999 & 10 & 33 & 3 \\
                                & \$1,000 - \$29,999 & 19 & 57 & 9 \\
                                & \$29,999 - \$74,999 & 5 & 13 & 2 \\
                                & >= \$75,000 & 10 & 12 & 3 \\
\bottomrule
\end{tabular}}
\caption{\change{}{Participant demographics and experience across studies. Questions were similar between studies but not exactly same.
~~FL: free-listing. FR: factor-rating. I: interview study.}}
\label{tab:demo}
\end{table}

\change{}{\autoref{tab:demo} summarizes participants' self-reported demographics and experiences.} We had 
\change{56 and 159 participants in the \firststudy and \secondstudy, respectively}{56 participants in the \firststudy, 159 in the \secondstudy, and 24 interviewees}. 
Participants were mainly from North America, South Asia, and Europe; young in age; and overwhelmingly male\change{(see~\autoref{tab:demo})}{}.  

Exact demographics of \hunters are unknown; however, \change{}{aside lower education levels,}{} our participants are similar to \change{\hunter demographics}{those} reported by popular \bbplatforms. \change{}{They report samples consisting of 77-85\% $<$ 35 years old, 90-96\% male, 19-34\% with graduate degrees, and 37-67\% working < 10 hours a day on \bbs}~\cite{hackerone2020report,bugcrowd2020report, bugcrowd2021report}. \change{}{HackerOne's sample---the only marketing survey to report \hunter experience---had similar levels of experience (i.e., 47\% $<$  three years of hunting) to ours~\cite{hackerone2020report}.}

\section{Factors \change{(RQ1, RQ2)}{}}
\label{sec:factors}

We identified 54 factors affecting participation in the \bb ecosystem under four \hlqs. \change{}{A complete list can be found in~\autoref{tab:refined_codes}.}
\change{}{Some of these factors are reported in prior work; 
as shown in \autoref{tab:refined_codes}, no previous work 
has systematically identified all factors, and most focus on 
benefits, leaving gaps in challenges and in which 
platform features are most useful. Further, we provide the first 
in-depth ranking of factor importance. 
}

\definecolor{Gray}{gray}{0.95}

\begin{table*}[t!] %
  \centering
  \footnotesize
  \begin{tabular}{l>{\raggedleft}p{1.68in}p{3.87in} p{0.1in} p{0.31in}}
  \toprule
  \midrule
  \textbf{} & \textbf{Factor} & \textbf{Definition} {\hfill} \textbf{\change{}{Relevant work}} & \textbf{FL} &  \change{}{ Worth\textbf{$(\pi)$}} \\
  \midrule
  \multirow{15}{*}{\normalsize\textbf{\rotatebox[origin=c]{90}{Choosing a program}}} 
        & \textbf{Scope} & Domains or assets included in the program.{\hfill} {\hfill}~\cite{bugcrowd2021report} & 28 & \change{}{ 0.156} \\
        &  \textbf{Reward} & Expected monetary or non-monetary rewards.  {\hfill}~\change{}{\cite{hackerone2020report,bugcrowd2021report, finifter2013empirical, zhao2015empirical, laszka2018rules}} & 36 & 0.120 \\
        & \textbf{Bounty table} & Reward rules and ranges for different bug types.{\hfill}\change{}{\cite{finifter2013empirical}} & 16 & 0.117 \\
        &  \textbf{Technology familiarity} & Familiarity with the technology of the assets (e.g., familiarity with web or iOS). {\hfill}~\change{}{\cite{bugcrowd2021report}} & 22 & 0.113\\
        & \textbf{Legal safe harbor} &  Includes a commitment to not pursue legal actions after hackers who follow the rules.~\change{}{\cite{ellis2022bounty}} & 4 & 0.098 \\
        &  \textbf{Program repute} & Reputation in the community for being pleasant to work with.{\hfill}~\change{}{\cite{hackerone2020report}} & 15 & 0.086\\
        & \textbf{Learning opportunity} & Lack of familiarity with the technology of the assets and interest in learning.{\hfill}~\change{}{\cite{hackerone2020report}} & 1 & 0.064 \\
        &  \textbf{Private or public} & Private programs (only by invitation) vs. public programs (accessible by anyone).{\hfill}~\change{}{\cite{hackerone2020report}} & 4 & 0.048 \\
        & \textbf{Company familiarity} & Company behind the program is widely known; you or your peers use its products.~\change{}{\cite{hackerone2020report, zhao2015empirical}} & 15 & 0.047 \\
        &   \textbf{Saturation} & Number of reports received or number of hackers working on the program. & 8 & 0.047 \\
        & \textbf{Career opportunities} & Future career opportunities with the company behind the program.{\hfill}~\change{}{\cite{hackerone2020report}} & 3 & 0.027 \\
        &  \textbf{Public disclosure} & Public vulnerability disclosure is generally allowed following the bug resolution. & 6 & 0.027 \\ 
        & \textbf{Age} & For how long the program has been running.{\hfill}~\change{}{\cite{maillart2017given, walshe2020empirical, sridhar2021hacking}}  & 6 & 0.023 \\
        &  \textbf{Business domain} & Business domain of the company behind the program (e.g., healthcare, retail).{\hfill}~\change{}{\cite{zhao2015empirical, sridhar2021hacking}} & 2 & 0.021 \\

      &  \textbf{Country} & Where the company behind the program is located. & 1 & 0.006\\

\midrule

    \multirow{17}{*}{\normalsize\textbf{\rotatebox[origin=c]{90}{Challenges of bug hunting}}} &
    
      \textbf{Poor responsiveness} & Lack of responses or slow responses from program managers.{\hfill}~\change{}{\cite{hackerone2021report, bugcrowd2020report,bugcrowd2021report, finifter2013empirical}} & 30 & 0.130 \\  
    &    \textbf{Dissatisfaction with responses}& Rewards are lower than expected (e.g., downgraded severity, impact).{\hfill}~\change{}{\cite{ellis2022bounty}} & 26 & 0.120\\
    & \textbf{Unclear scope} &  Program scope is not defined clearly. & 3 & 0.082\\
    &  \textbf{Poor platform support}& Dissatisfaction with how platforms handle issues, such as mediations. & 1   & 0.079   \\
    & \textbf{Duplicates}& Too many reports marked as duplicates.{\hfill}~\change{}{\cite{ellis2022bounty}} & 4 & 0.078  \\
    &  \textbf{Assets outside expertise} &  Assets are outside area of expertise, lacking certain required skills.&  11 & 0.068  \\
    & \textbf{Secure assets} & Finding bugs is too difficult. & 5 & 0.064\\
    &  \textbf{Stress and uncertainty} & Fear of burning out, social isolation during work, irregular income, etc.{\hfill}~\change{}{\cite{ellis2022bounty}} & 5  & 0.062 \\
    & \textbf{Too much labor work} & Menial tasks (e.g., CAPTCHA, timeouts, obfuscation, setting up test accounts). & 12 & 0.050\\
    &  \textbf{Boredom}& Bored of working on the program or a more interesting program launches.{\hfill}~\cite{ellis2022bounty}& 8 & 0.050\\
    & \textbf{Unrepresentative reputation system} & Hackers' reputation points do not reflect real experience.  & 1 & 0.047\\
    &  \textbf{Difficulty working with managers} & Managers are difficult to work with (e.g., disrespectful, requiring extra work).{\hfill}~\change{}{\cite{hackerone2021report,ellis2022bounty}} & 23 & 0.044\\
    & \textbf{Not enough time}& Not having enough time for participating in bug bounties. & 2 & 0.043 \\
    &    \textbf{Limited vulnerability disclosure}& Restrictive vuln. disclosure policies and NDAs that may prevent you from publishing.~\change{}{\cite{ellis2022bounty}} & 2 & 0.041\\
    & \textbf{Legal threats}& Fear of threats of legal implication (civil or criminal).  & 2 & 0.028 \\
    &   \textbf{Communication or language} & Communication difficulties from lack of language skills, anxiety in communication, etc. & 2 & 0.014\\

\midrule
    
     \multirow{10}{*}{\normalsize\textbf{\rotatebox[origin=c]{90}{Benefits of bug hunting}}}
     &  \textbf{Monetary rewards} & Monetary compensation.{\hfill}~\change{}{\cite{ellis2022bounty,hackerone2019report,hackerone2020report,hackerone2021report,bugcrowd2020report,bugcrowd2021report, zhao2015empirical}} & 42 & 0.191 \\
     &   \textbf{Learning} & Learning or improving skills.{\hfill}~\change{}{\cite{ellis2022bounty,hackerone2019report,hackerone2020report,hackerone2021report,bugcrowd2020report,bugcrowd2021report}} & 32 & 0.170 \\
     &  \textbf{Enjoyment}& Enjoyment or challenge of white-hat hacking.{\hfill}~\change{}{\cite{ellis2022bounty,hackerone2019report,hackerone2020report, hackerone2021report,bugcrowd2021report}} & 20 & 0.140\\
     &    \textbf{Legal safe harbor}& Hacking without the threat of legal actions if they obey the rules. & 4 & 0.118\\
     &  \textbf{Flexibility}& Work schedule and place flexibility (compared to traditional employment).{\hfill}~\change{}{\cite{ellis2022bounty, bugcrowd2020report}} & 16 & 0.095 \\
     &   \textbf{Career}& Building relations with companies for employment and other opportunities.~\change{}{\cite{ellis2022bounty,hackerone2019report,hackerone2020report, hackerone2021report}} & 11 & 0.091\\
     &  \textbf{Community}& Bug bounty creates a community of hackers.{\hfill}~\change{}{\cite{ellis2022bounty,hackerone2020report}} & 3 & 0.071 \\
     &   \textbf{Altruism}& Improving cybersecurity to help others, securing the internet.{\hfill}~\change{}{\cite{ellis2022bounty,hackerone2019report,hackerone2020report, hackerone2021report, bugcrowd2020report,bugcrowd2021report}} & 5 & 0.062\\
     &  \textbf{Reputation}& Earning platform reputation points, building a following, etc.{\hfill}~\change{}{\cite{hackerone2019report,hackerone2020report,ellis2022bounty,walshe2020empirical}} & 14 & 0.048 \\
     &   \textbf{Non-monetary rewards}& Non-monetary compensation (e.g., SWAG, hardware, subscriptions).{\hfill}~\change{}{\cite{ellis2022bounty}} & 4 & 0.013 \\[0.5em]

\midrule

    \multirow{13}{*}{\normalsize\textbf{\rotatebox[origin=c]{90}{Useful platform features}}}
    &   \textbf{Ease of payment} & Receiving payments in a standardized, hassle-free way. & 12 & 0.156 \\
    &     \textbf{Ease of reporting} & Easy to generate, submit, and track reports and their status. & 16 & 0.142 \\
    &   \textbf{Viewing disclosed vulnerabilities} & Platform-provided interface for viewing bugs found by others.{\hfill}~\change{}{\cite{zhao2015empirical}} & 15 & 0.137 \\
    &     \textbf{Private program invitations} & Access to private programs on the platform. & 5 & 0.107 \\
    &   \textbf{Program directory} & Listing many programs in one place, with statistics, details, etc.  & 17 & 0.068 \\
    &     \textbf{Standardized rules} & Platform standardizing how scopes, rewards, criticality, etc. are defined. & 3 & 0.063 \\
    &   \textbf{Community} & Platform making effort to create a community of hackers.{\hfill}~\change{}{\cite{ellis2022bounty}} & 11 & 0.057 \\

    &   \textbf{Platform rewards} & For example, platform SWAG and funded travel.{\hfill}~\change{}{\cite{ellis2022bounty}} & 1 & 0.054 \\
    &     \textbf{Mediation} & Platform resolving disputes between hackers and programs. & 13 & 0.051  \\

    &     \textbf{Platform-managed disclosure} & Platform-provided tools/mechanisms to publicly disclose resolved bugs. & 6 & 0.050 \\
    &   \textbf{Resources for learning} & Platform providing free resources on how to hack (e.g., Bugcrowd University).{\hfill}~\change{}{\cite{hackerone2021report}} & 2 & 0.047 \\

    &     \textbf{Reputation system} & Platform managed-reputation system for hackers. & 6 & 0.043 \\
    &   \textbf{Platform triage} & Triaging managed by the platform (e.g., HackerOne triages your report instead of Uber). & 5 & 0.023 \\
\midrule
\bottomrule

 \end{tabular}

\caption{\change{}{Factors used in the \secondstudy, organized by  \hlqs. FL: count of participants who listed the factor in the \firststudy. Worth ($\pi$): the estimated relative importance of factors (values only meaningful in comparison, see~\autoref{subsec:ratingmethods} for details). Descriptions shortened for space; full versions are given  in~\autoref{app:full_list}. Citations show factors identified in prior work. %
}}
\label{tab:refined_codes}
\end{table*}

In our interviews, we identified seven common themes that 
connect interrelated factors; we use these themes to 
organize our discussion. For brevity, we primarily discuss the most popular factors; details about more can be found in\ifIsArxiv~\autoref{app:additional_factors}\else~the extended paper~\cite{akgul2023bug}\fi.
We report how many interview participants made an argument (I), how many \firststudyps listed a factor (FL), and \change{the mean rank (out of seven) obtained in the}{the relative importance of factors (worth estimates) generated from the} \secondstudy
\change{($\mu_r$)}{$(\pi)$.\footnote{\change{}{Worth estimates can only be meaningfully compared within \hlqs, which were analyzed separately; \autoref{tab:refined_codes} provides a complete list.}}} 

It is important to note that because each interviewee only had time to comment on a subset of factors, participant counts from interviews are provided for context but cannot be interpreted as prevalence.

\subsection{Earning rewards}
\label{sec:rewards}
Though rewards might seem like an obvious motivation~\change{}{
\cite{ellis2022bounty,hackerone2019report,hackerone2020report,hackerone2021report,bugcrowd2020report,bugcrowd2021report, zhao2015empirical}}, \hunters expressed nuanced details.

\paragraph{Reward considerations} Unsurprisingly, \textit{monetary rewards} were commonly listed by \firststudyps and were highly ranked by \secondstudyps both as a consideration for choosing a \bbprogram (\FL{36}, \change{\mean{5.91}}{\w0.120}) and as the top benefit of \bbs (\FL{42}, \change{ \mean{6.31}}{\w0.191}). \change{Further, }{}\Hunters similarly prioritized the mapping of bug severity to reward (i.e., the \textit{bounty table}, \FL{16}, \change{\mean{5.87} }{\w0.117}). \textit{Non-monetary rewards} (e.g., swag\change{free travel}{}; \FL{4}, \change{\mean{4.35}}{\w0.013}) were \change{considered less important when choosing between \bbprograms. }{motivators~\cite{ellis2022bounty}, but were ranked least important.}

\change{While the frequent mention of monetary rewards is not especially surprising, 
}{}In interviews, \hunters described nuanced preferences
for how bounties are determined and managed.
Many \hunters (\I{9}) argued that bounties should be correlated with the 
severity of the identified bug; this aligns well with industry practices~\cite{applebb}. Three specifically mentioned valuing 
bugs based on how much they might cost the company if exploited. Three 
\hunters who specialized in high-severity bugs mentioned the importance of large 
rewards for these bugs, with less concern about rewards for less critical  
bugs. \change{As one noted, ``For low severity \ldots~I do not mind getting rewards or not getting the rewards.'' }{}

In contrast, three \hunters primarily considered the payout amounts 
for low- and medium-severity bugs when choosing a \bbprogram, mainly 
because they considered finding more critical bugs improbable. One 
\hunter said that they would only take high-severity payment levels into 
account if they planned to commit to a program for a long time. 
Other \hunters (\I{7}) argued that payments should be proportional to \hunter 
effort, not just the outcome. %

Some (\I{7}) argued the bounty table was not a good predictor of their 
expected payment for participating, which would be determined largely by factors such as how many bugs they expected to be able to find, possible bonus payments (e.g., for well-written 
reports), and the potential for multiple payments (e.g., for re-exploiting 
a previously \change{resolved}{patched} bug). One noted a preference for a program that 
``doesn't have really huge payouts, but \ldots~you're likely to get more 
out of it because you actually submit more bugs because you know it better.''

Interestingly, eight \interviewps indicated there were other, more important motivations than monetary rewards. Generally motivated by finding as many bugs as possible, one \hunter noted, ``I try sometimes finding such programs, who doesn't give monetary rewards. \ldots It's easy to find vulnerabilities from such programs.'' Another said they report bugs \change{even if they will not get paid}{regardless of payment}, because ``\change{I don't care about the money.}{}I just want to make companies secure.''

In lieu of monetary rewards, \change{}{as noted in prior work~\cite{ellis2022bounty},} a few \interviewps were content to receive only a reputation boost (\I{3}); more (\I{5}) found merchandise sufficient compensation. \change{}{However, \secondstudyps overwhelmingly rated reputation and non-monetary rewards as the least important motivators.}

\subsection{Learning opportunities}
\label{subsec:learning}

All three studies highlight the importance \hunters place on being able to learn. \Hunters report learning in many ways, e.g., from publicly disclosed bugs and community interactions. 

\paragraph{Importance of learning}
\textit{Learning} new hacking techniques was one of the most frequently listed benefits in the \firststudy and was found to be the second most important benefit by the \secondstudyps (\FL{32}, \change{\mean{6.18}}{\w{0.170}}). Though not as immediately apparent, \hunters' focus on learning was evident in several other, related factors that were commonly discussed.

Referring to the learning opportunity that \bbprograms provide, most \interviewps said that learning is an integral part of the process: a \hunter will either learn organically or be forced to learn new techniques to stay relevant (\I{12}). Specifically, three explained the need to learn new technologies and methods when going after certain bugs. One hunter recalled having to learn GraphQL to find a bug: ``What I thought was, let's learn about it first. So I got into learning GraphQL, and the structures, how it's written, how you get response, what kind of response you get.'' %

\change{Practicing is integral to mastering new skills~\cite{votipka2021hacked}.}{} Three \hunters noted that \bbs also provide an opportunity to practice skills, \change{contributing to learning.}{an integral part of learning~\cite{votipka2021hacked}.}

In contrast, three \hunters noted that the constant pressure to learn can be difficult. %
``No matter how experienced you are, everyday you will need to learn new things. And I think this is one of the reasons that makes bug hunting a bit difficult, or a bit tricky. That if you stop learning, you will lose the [touch]. In order to find bugs, you will have to stay updated in community to learn something new.''

\paragraph{Learning from public reports}
Our \interviewps \change{highlighted}{confirmed previous speculation~\cite{zhao2015empirical}} about the importance of \textit{public disclosure} to the learning process. 
With learning as a top benefit, \hunters naturally valued 
public disclosure as a learning opportunity. 
While \secondstudyps did not directly value a program's willingness to disclose as much as other factors (\FL{6}, \change{ \mean{4.63} }{\w0.027}), they did consider viewing disclosed vulnerabilities to be the platforms' third most useful feature (\FL{15}, \change{\mean{6.41}}{\w0.137}). 
This result adds nuance to prior work~\cite{ellis2018empbounty}, 
suggesting that even though \hunters do not need a specific \bbprogram to be disclosure friendly to work on it, they prefer the entire ecosystem to be so. \change{}{This reinforces the notion that \hunters primarily seek public disclosure as a resource for learning, with reputation building less important.} Conversely, \textit{resources for learning} provided by \bbplatforms were not rated as useful (\FL{2}, \change{\mean{5.59}}{ \w{0.047}}), suggesting room for improvement.

Review of publicly disclosed bugs was \hunters' most frequently mentioned learning method (\I{18}). These reports show \hunters how their peers approached the problem (\I{9}): ``It's all about the thought process.'' Reported bugs can also potentially be directly reproduced in other programs (\I{4}). A \hunter described this as, ``everyone will take the exploit or the attack vector from the report, and everyone will be trying to score the same issue on different programs.''
\change{Some \interviewps (\I{5}) mentioned finding independent write-ups (e.g., blog posts) by other \hunters. However, more \hunters mentioned disclosures made through \bbplatforms (\I{11}).}{}

Public disclosure is not only about learning new hacking techniques; \hunters noted that disclosure also helped with evaluating \bbprograms (\I{10}), particularly 
the quality of the triage team (\I{3}) (see~\autoref{subsec:disputes}). As one \hunter put it, ``I read about all the disclosed reports on HackerOne and I see how they communicate with the researcher.'' %
Further, two \hunters viewed public disclosures as advertising opportunities for programs. The disclosures grabbed their attention as they skimmed platforms' disclosed vulnerabilities lists (e.g., Hacktivity and Crowdstream~\cite{hacktivity,crowdstream}). %
These disclosures tell \hunters that a company is taking its \bbprogram seriously, making it a more attractive organization to work with.

\paragraph{Impact of community}
\Firststudyps reported the \textit{community} of \hunters as a benefit of \bbs (\FL{3}). Though not \change{ranked}{rated} as highly as other factors \change{factors in the \secondstudy}{} (\change{\mean{5.52}}{\w0.071}), we find the community to be an integral part of learning. This community develops naturally through interactions on social media, but many \hunters (\FL{11}, \change{\mean{5.77}}{\w0.057}) also noted \bbplatforms' contributions (e.g., live hacking events).

Nearly all \interviewps noted that the \hunter community frequently interacts with each other and shares information, creating a learning environment (\I{22}).  %
This includes \hunters disclosing as a way of ``giving back to the community'' (\I{5}), providing specific technical knowledge (\I{7}), and answering questions ($n$=5). One \hunter appreciated the responsiveness of blog-post authors: %
``If you have any doubt, you just ping them. \ldots The people are very helpful. The communities are good.'' Sharing of resources---commonly free of charge (\I{1})---is particularly striking in the competitive world of \bbs (i.e., only one \hunter gets paid for each vulnerability) (\I{2}). %

The bug-bounty community also offers the ability to learn by developing relationships with other \hunters. %
Many participants mentioned gaining professional contacts that were helpful to their career (\I{8}), and five of those \change{specifically noted that these contacts produced direct collaboration on bug-finding efforts}{noted these contacts led to bug-finding collaborations}. Several also highlighted the social benefits of community engagement (\I{7}), which can reduce the isolation of working remotely 
and individually (\I{2}). 
In-person events in particular offered this sense of belonging. One \hunter explained, ``you get to meet so many people. You tend to have so many opportunities, and people seem to recognize you, and everyone wants to be your friend, you want to be everyone's friend.'' \change{And everyone is so friendly in this community.''}{}%

Not everyone saw the community as a useful learning tool. Five \interviewps noted that community members do not always share useful knowledge, 
and two said sharing is a relatively recent phenomenon. One noted that ``there are a few too many trolls, but usually they're [the average \hunter{}] quite helpful and that does really encourage things.''

\paragraph{Contrast to technology familiarity} \Firststudyps noted that \textit{technology familiarity} was important when picking \bbprograms (\FL{22}) and factor-ranking participants ranked it an important factor (\change{\mean{5.86}}{\w0.113}). 

Most \interviewps implied that they preferred programs with technologies they are familiar with because it allows them find bugs more easily ($n$=15). Specifically, some ($n$=5) noted that familiarity with the technology \change{}{(sometimes referring to specific libraries, ``you're familiar for instance with Django, probably with Flask or things like that, so you just go and search if the version is updated.'')} helps them apply skills they already have.
As one put it, ``\ldots \change{so the first point is that}{} if you want to find any bugs, you will need to be familiar with the technology\change{, what the developer is using.}{\ldots}'' Three \hunters said that familiarity helps them with finding higher severity bugs, while one said it helped them with identifying ``the low hanging fruit.'' %

On the surface this might appear to contradict a focus on learning by emphasizing existing skills. However, interviews suggest otherwise.
Participants reported that even existing skills need to be practiced (\I{3})~\cite{votipka2021hacked} and \hunters learn new techniques regardless of the target. As such, we argue that \hunters picking more familiar technologies does not necessarily limit their learning. Rather, focusing on familiar technologies allows \hunters to get started on finding bugs and offers opportunities to learn about new variations or connected modules and to master skills through practice. %

\subsection{Predicting the likelihood of finding a bug}
\label{sec:finding_bugs}

Many participants noted factors related to the probability of finding 
a bug as important considerations. These factors include program scope, age and saturation, whether the program is public or private, the likelihood of duplicates, and the impact of the hunter's reputation. There was little consensus as to what aspects of a given factor were 
beneficial or not. %

\paragraph{Extent and clarity of program scope} Half of \firststudyps
identified the \textit{scope} --- which dictates which assets can be investigated and what bug types will be accepted --- to be important for choosing a \bbprogram (\FL{28}). 
\Secondstudyps ranked scope the most important factor (\change{\mean{6.10}}{\w0.156}). \change{}{Interestingly, this finding has only been previously mentioned in marketing materials~\cite{bugcrowd2021report}.}

Many \hunters consider scope to be a useful predictor of how many bugs they can expect to find. Most \interviewps preferred larger scopes (\I{15}), reasoning that a larger attack surface should correlate with more potential bugs. For example, one \hunter explained, ``a good example of a big scope is the U.S. Department [of Defense], \ldots 
you have a lot of systems where you are allowed to identify vulnerabilities, which gives you a lot of possibilities to identify things.'' 

Larger scopes also allow \hunters to evaluate a product as a whole (as a malicious actor could); two said better understanding of the overall architecture of assets ultimately creates a higher chance of finding bugs. As one explained, ``I just want to find out if some adversary can get their hands on customer data. And they [adversaries] don't have scopes, they just hit the target, and I want to do the same.'' 

Correspondingly, narrow scopes have important drawbacks (\I{5}). First, they raise 
the risk of finding out-of-scope bugs (\I{3}). Further, one argued that \bbprograms cannot sufficiently isolate small scopes within highly connected products: %
``They have this big list of out-of-scope,\ldots. So I go hit their site and I go to an in-scope target and it starts hitting all of these out-of-scope targets all over the place.'' 
Narrow scopes also increase the chances of \bbprograms becoming saturated (defined below), because there are likely more \hunters per target (\I{3}), and the difficulty of finding a bug increases accordingly 
(\I{2}). %

Conversely, a few did not consider a wide scope to be a major factor. Two \interviewps noted scope was more important to beginners compared to the experienced, \change{who may have the skills to identify harder bugs in limited scopes}{due to increased skills}, ``Maybe it's less true today because I'm more experienced now, but at the beginning at least it was important.''

Participants considered \emph{unclear scopes} to be a major issue with \bbs (\FL{3}, \change{\mean{4.99}}{\w0.082}). Many \interviewps recalled frustrating experiences with imprecise scopes (\I{10}), including those that were vague (\I{5}) or outdated (\I{2}). Two noted that they had to talk to managers for clarification. \change{}{Unclear scopes are in many cases a type of communication issue; we discuss major challenges related to communication in~\autoref{subsec:disputes}.}

\paragraph{Age and saturation}
Participants said program age---time elapsed since launch---affected 
which \bbprogram they would select (\FL{6}, \change{\mean{4.44}}{\w0.023}). A closely related factor 
is saturation: how much attention a program has already received and how many 
\hunters are actively working on it, relative to the scope (\FL{8}, \change{\mean{5.17} }{\w0.047}). 
Eleven participants use age as a proxy for saturation, generally expecting younger programs to have more remaining bugs. 
As such, 12 \interviewps prefer newer programs.
Older programs do roll out new code periodically (\I{3}), creating 
new opportunities.

\change{On the other hand}{However, there was no consensus on this point. Seemingly contradicting prior work \cite{maillart2017given, walshe2020empirical},} six preferred older programs, due to more experience handling bug reports (\I{2}) and provide more publicly available information (\I{1}). One participant said, ``There's going to be fewer reports about that [younger] 
program, so you're not going to learn as much.''

\paragraph{Public vs. private programs}
A \textit{public} program is openly advertised, and anyone can participate. In contrast, \textit{private} programs 
allow participation by invitation only. Several \interviewps used public/private status to judge the chances of finding a bug when choosing programs; however it was not ranked as a top factor (\FL{4}, \change{\mean{5.16}}{\w0.048} ). \Interviewps' preferences were evenly divided (private: \I{11}; public: \I{9}). 

Private programs were reported to have fewer hunters (\I{9}), suggesting lower saturation. 
Conversely, three said that 
private programs are just as saturated as public: ``Most programs tend to have a pretty healthy amount of hackers regardless.'' Further, three noted that public programs often have wider scopes.
Aside from saturation, some used public/private status as a heuristic for age (\I{1}), responsiveness (\I{1}), and possibility of public disclosure (\I{2}).

\paragraph{Likelihood of duplicates}
Reporting a \textit{duplicate}---a previously reported bug---frequently goes unrewarded, creating frustration for \hunters. This was ranked as the fourth largest issue (\FL{4}, \change{\mean{5.02}}{\w0.078}). \Interviewps associated duplicates with 
\managers not patching bugs soon enough (\I{6}) 
and older programs (\I{2}): ``If you're among the first ones on the program, 
I think that \ldots~increases your chances not to hit a duplicate.'' Three argued that 
duplicates are primarily a problem when \hunters go after easier to find bugs, and can reduced by avoiding ``the low hanging fruit."%

Some \interviewps (\I{4}) argued that duplicates are inevitable: ``Duplicates are always the thing that you just have to get used to.'' However, several (\I{9}) described coping mechanisms: three said it helps knowing the bug they found has been acknowledged, four said being added to the report (even without rewards) provides assurance the \managers are not lying, and two said they would not go after bugs they consider popular among the community.

\paragraph{Impact of reputation}
One factor in whether a \hunter will receive rewards is the \hunter's \textit{reputation}. 
\change{Though reputation was mentioned by 14 participants in the \firststudy as a benefit, it was ranked as the second-least important benefit (\change{\mean{5.36}}{\w0.048}).  We speculate that although reputation is a very visible aspect of bug bounties
(often how \hunters are \change{compared and}{} ranked\change{, visible in \hunter profiles}{}), 
it's ultimately not as beneficial as other factors. Our results indicate this might not hold for \lowskill \hunters (also see~\autoref{subsec:skill_levels_comp}).}{}
Reputation can be built through official platform reputation systems, as well as through basic publicity of found bugs with write-ups.
\change{Nine \interviewps}{\Hunters with both high and low reputation (\I{9}) said they 
will not be or are not taken seriously when reporting bugs for lack of reputation;}
however, reputation is typically acquired by 
reporting valid bugs, making it hard for beginners to get started. Similarly, five noted that \hunters{}' reputations help them garner private invites, `to make sure that I stay up high enough for them to keep going and giving me private invitations.'' \change{Otherwise, you start losing them and that's not really great.}{} Four participants mentioned that reputation helps with being taken seriously by other security researchers \change{}{as well} (\I{3}) \change{or by the \managers (\I{3})}.

Though 14 \firststudyps mentioned reputation as an overall benefit, \secondstudyps gave it limited importance, rating it the second least important benefit ({\w0.048}). Similarly, the \emph{reputation systems} that \bbplatforms provide were rated the second least important feature (\FL{6}, \w{5}). We speculate that although reputation is a very visible aspect of bug bounties, 
it's ultimately not as beneficial as other factors. This finding contradicts prior work that hypothesizes 
reputation might be more important than or comparable to monetary compensation~\cite{ellis2018empbounty, walshe2020empirical, sridhar2021hacking}.

\subsection{Communication, disputes, and mediation}
\label{subsec:disputes}
The \firststudy revealed multiple factors relating to 
communication between \hunters and \managers, including
poor responsiveness, rude or unreasonable \managers, and unexpected \change{outcomes of submitted}{responses to} reports. \change{Some of which has already been identified in prior work}{Though some of these issues have been discussed in prior work (see~\autoref{tab:refined_codes}), other important issues, such as the mediation process, were not previously explored}.

\paragraph{Poor responsiveness} In the \firststudy, a majority of participants (\FL{30}) mentioned \textit{poor responsiveness}: where \managers do not efficiently communicate with \hunters. Responsiveness was also ranked as the top challenge in \bbs by \secondstudyps (\change{\mean{5.39}}{\w0.130}). 
\change{Though a majority of \interviewps also mentioned poor responsiveness as an issue (\I{17}), we found no consensus on what constitutes a timely response.}{}
\change{}{Prior work~\cite{ellis2022bounty} and marketing materials~\cite{bugcrowd2021report} have briefly acknowledged the importance of this issue. Our \interviewps (\I{17}) were able to elaborate on what constitutes a timely response, including time to first response (\I{1}), time to severity determination (\I{1}), time to payout (\I{5}), and/or time to produce a patch (\I{1}). This diverse list of definitions suggests \bbprograms should work to improve all metrics, rather than just picking one.}

Some (\I{5}) recounted frustrating instances where the \managers would not respond to their or others' communication attempts at all.
One said, ``Sometimes some programs don't fix the issues, which are valid, and close them as not applicable, with no reason. Once I comment on it, they don't respond on it and totally ignore my comment.'' %

Many \interviewps suggested ways to make unresponsiveness more tolerable, but 
with little consensus. 
Some \hunters wanted frequent updates for payout delays (\I{4}), which are ``acceptable, as long as they are giving updates.''\change{But if they are sitting silent, no updates, \ldots then the program is a no-no for me.}{} Two said they could wait longer for higher severity bugs; one, conversely, expected faster resolution for higher-severity bugs. 
\change{Though tough to implement by \managers, a few \hunters were more tolerant of long payout times when they themselves were financially stable (\I{3}).}{}

\paragraph{Difficulty working with managers/triagers}
Though responsiveness was often the most frustrating communication problem, it was not the only one. Notably, \textit{difficulty working with managers} was mentioned by many \firststudyps (\FL{23}).
However, this factor ranked among the least challenging (\change{\mean{4.48}}{\w{0.044}}), possibly because we distinguished it from the closely related factor of dissatisfaction with responses (discussed next).

Six \interviewps mentioned \managers are not always professional in their communications; they can be rude (\I{3}) or otherwise dismissive (\I{2}). 
As one \hunter said, 
``It's actually not fun to work with somebody who insults you or is rude or stuff like this.'' %

While most \hunters attributed communication difficulties to program managers, 
nine \interviewps blamed \hunters too. Six said bug reports are sometimes not clear. In response, one \hunter was trying to improve: ``Something that we've also been trying to do a lot now is \ldots %
try to make the reporting as quality and clear as possible.'' %
Two \interviewps said \hunters are occasionally rude and should behave professionally ($n$=2). %

Another source of potential conflict is \emph{platform triagers}, 
mentioned by five \firststudyps but ranked as the least useful 
platform feature (\change{\mean{5.06}}{\w{0.023}}).
Some \bbprograms hire triagers from \bbplatforms, rather 
than maintaining their own team. 
A few \hunters (\I{2}) argued that triagers outside the development 
team increase communication difficulty. Two said for-hire triagers create 
an unnecessary barrier between the \hunter and the team that will 
fix the bug; another noted that since the company makes the final payout 
decision, triagers' severity gradings are irrelevant. 

Conversely, five \interviewps had positive experiences with platform triagers. They noted that platform triagers specialize in dealing with \hunters, and therefore are better to work with than \managers. One appreciated that platform triagers help clarify bug reports before they get to \managers, and another said platform triagers helped them trust decisions they disagreed with (e.g., duplicates). Two participants recalled negative experiences with programs that do not have platform triagers. %

When asked about dissatisfaction with managers, two \hunters explicitly said they had no major issues; one said, %
``For the most part, I've had a really positive experience.'' %

\paragraph{Unexpected responses}
Closely related to the previous factor, \textit{dissatisfaction with responses} was a commonly mentioned, highly frustrating problem (\FL{26}, \change{\mean{5.36}}{\w{0.120}}).

Most \interviewps recalled disagreeing with the \managers' evaluation of a bug (\I{16}), usually about vulnerability applicability and severity levels, duplicate status, and permission for public disclosure. 

Eight said \managers make errors related to misunderstanding  
submitted reports, either because they don't read them thoroughly (\I{2}) or don't understand technical details (\I{4}). Conversely, one noted that \hunters might not fully read program descriptions, ``To be honest, I sometimes forget to read the whole program scope and sometimes I submit some vulnerabilities that are not in scope.'' %

Troublingly, as found before \change{}{in similar contexts}~\cite{alomar2020nicebugs}, two \hunters noted they may not report potential bugs to avoid scope disagreements with \managers. 

Five \hunters were more pessimistic, arguing that companies might trick \hunters in order to avoid paying them. One \change{\interviewp said}{recalled}, ``Sometimes, when the program got everything they needed from you, \ldots they set the status, [to] needs more \ldots Then, the bug is getting fixed \ldots severity gets lowered, and you get lower rewards than you expected. And then, they just suddenly stop responding to your comments, and that's a situation that happens all the time.''

Disputes and responsiveness issues can, in rare situations, lead to extreme outcomes. One participant was banned from a major \bbplatform. Another participant noted that exploiting the bugs themselves becomes more attractive when they get frustrated with \bbprograms.

\paragraph{Attempting to resolve disputes} 
When disputes arise, \hunters may enlist the \bbplatform as an 
ostensibly impartial mediator, although they generally cannot override the \managers' final assessment~\cite{hackeroneMediation}.
Although listed by many in the \firststudy as a benefit (\FL{13}), \textit{mediation} was not ranked a top feature of \bbplatforms (\change{\mean{5.77}}{\w0.051}). Similarly, \textit{poor platform support} (including for mediation), was seen as one of the biggest challenges faced in \bbs (\FL{1}, \change{\mean{5.03}}{\w0.079}).

\Interviewps were split 
between negative (\I{8}) and positive (\I{5}) reactions to platform 
mediation. On the positive side, some noted that mediation was useful to clarify technical issues to the \managers (\I{5}). One explained, ``One of the biggest benefits %
\ldots is they do facilitate that conversation. %
\ldots Mostly to help a non-technical program manager understand the impact.'' A participant found comfort in knowing \change{they could ask for help, ``It's always very soothing to know that the platform has your back. In most cases.''}{``\ldots the platform has their back. In most cases.''}%

Other participants, however, argued that mediation is inherently biased,
because \bbplatforms favor the companies that pay them to host (\I{4}). 
Two recalled experiences where \bbplatforms did not respond to mediation requests; 
two others had heard of but not experienced such issues. %
One \hunter described requesting mediation, ``and nothing happens. You get a reply every two weeks because that's how often you can ping them, and there's no indication that there's anything that happens. It is the biggest joke.''

\subsection{Gig-work benefits and drawbacks }
\label{sec:freelance}
Multiple factors we identified relate to the gig-work nature of 
the \bb ecosystem. 
This model has benefits (e.g., flexibility) 
but also serious drawbacks, including stress and uncertainty. \change{}{
Notably, although they do not formally define and measure the prevalence of each issue, this aspect of \bbs has been heavily criticized by Ellis and Stevens~\cite{ellis2022bounty}}.

\paragraph{Flexibility}
A commonly referenced and middle-ranked (\FL{16}, \change{\mean{5.85}}{\w0.095}) benefit of the gig-work model is \textit{flexibility}: working from anywhere, at any time and for any duration, as well as full autonomy in what to work on.

Most \interviewps mentioned flexible work hours (\I{14}). Three emphasized that 
this avoids deadlines and pressure: ``Compared to my developer work that was 
all project based with deadlines \ldots here I don't have any deadlines.'' 
Others mentioned choosing what to hack (\I{2}), being able to work remotely (\I{2}), and the relatively low barrier to entry (not requiring an official position or prior approval, \I{1}). 

A few \hunters identified some drawbacks of flexibility. One noted that maintaining 
your own work hours can be tricky. Another mentioned that choosing when to work 
is less meaningful for those who hunt bugs essentially full-time: ``I can do whenever I want to. \ldots This could be harder if your only job is full-time bug bounty hunting.''

Further, \interviewps argued that the appearance of a low barrier to entry 
is deceptive: two said they had to build skills over years before they 
could participate effectively.

\paragraph{Negative aspects of gig-work}
\Bbs typically only reward \hunters for bugs deemed
unique and valid. As such, \hunters can spend significant effort for no pay, creating \textit{stress and uncertainty} (\FL{5}, \change{\mean{4.8}}{\w0.062}). This stress is in many cases downstream of  key factors we identified in prior subsections, such as unclear scope or likelihood of duplicates.

The most frequent source of uncertainty was payment (\I{9}). \Hunters do not know when they might find a bug, when and if \change{\managers might acknowledge it}{it might be acknowledged}, or when managers might decide to pay. Four participants noted the irregular income is especially stressful when \bbs are the main source of income: ``I also have days or even weeks where I don't find any single bug. And this is kind of depressing, I would say, which is the dark side of the whole story.''

Payment uncertainty is compounded by the competition to find bugs, ``because basically bug bounty is a competition, it's first come, first serve.'' Three \hunters noted that time-restricted events increase this sense of competition.

Even if a \hunter finds a bug first, they still must convince a \manager it is valid before they receive payment. As discussed in Section~\ref{subsec:disputes}, this can be  challenging for several reasons, creating additional stress and uncertainty.

\Interviewps described their approaches for coping with this uncertainty. 
\change{}{Confirming prior work~\cite{ellis2022bounty},} four said that for financial stability, they keep a full-time job outside 
the bug-bounty ecosystem. One said, ``I have a full-time job, and that makes my income a lot more stable. But I think if I was doing bug bounties full-time \ldots that might have impacted my mental health.''
Another three said they deal with the stress of \bbs by taking breaks: ``Whenever I feel burnt out, I just do something else. I give training \ldots or I go for a walk.\change{I have hobbies and stuff outside of service security.'' Notably, being able to afford breaks is a luxury not everyone has.}{''}

Five \interviewps said this uncertainty keeps them from doing \bbs full time. 
One specifically said even a potentially high income was not 
worth the instability: ``\change{I'm doing the side job thing purely because I'm not confident that I'll be able to live life entirely with a very secured income.}{}I have the potential of earning over 150 grand with bug bounty \ldots if I work at it, but the staggered amount that you get the money in is a problem. So if you're actually planning to do this full-time, you need to be able to have a bit of savings \ldots so you can push back on that in case.'' %
Conversely, one participant volunteered that they are a full-time hunter, and another argued that \bbs can be a valid career by themselves.

\subsection{Fundamental platform features}
\label{subsec:platform}

\Hunters listed multiple \bbplatform features as useful in the \firststudy. Most of these features exist independently of \bbplatforms and are discussed in previous sections. Here, we discuss the few that are unique to \bbplatforms and important to \hunters.

\paragraph{Ease of payment}
With monetary payouts a significant motivator, ease of payment was also of interest to \hunters (\FL{12}, \w{0.156}). The most frequently mentioned perk of \bbplatform payments were the many options provided (\I{7}). For example, six \interviewps preferred to be paid in an international currency and three preferred cryptocurrency payments. One explained, ``I prefer Bitcoins, because I have had problems with PayPal in the past.'' Some \hunters liked the simplicity of \bbplatform payments (\I{6}): ``I put in my checking account number one time and never think about it again.'' Finally, one enjoyed support for splitting bounties between collaborators and another liked that they could track all bounties earned.

\paragraph{Ease of reporting}
Standardized reporting, submission, and correspondence were appreciated by many \firststudyps (\FL{16}). In fact, this feature was rated the second most important a platform provides (\w{0.142}).

\Interviewps mostly mentioned the usefulness of the reporting interface (\I{9}) including guidelines for how to write reports, formatting tools (e.g., markdown), and support for visuals (e.g., videos). \Hunters also appreciated the interface for tracking bug progress (\I{5}), including correspondence with and actions taken by the \managers. 

A minority saw room for improvement (\I{6}), including larger size limits for reports and attachments, more consistency in reporting interfaces between different \bbplatforms, and a more formal communication process.

\paragraph{Program directories}
\textit{Program directories} are arguably the main feature of \bbplatforms, allowing \hunters to discover a variety of programs and compare several aspects of them in one place (\I{12}). 
Some of the available metrics reflect what \hunters care about the most (e.g., expected bounties, responsiveness). Accordingly, this feature was mentioned by most \firststudyps (\FL{17}) and rated relatively important by \secondstudyps (\w{0.068}).

A minority of \interviewps said there was room for improvement, such as the publication of more statistics (\I{3}) and \hunter feedback (\I{1}) for 
each program. %

\paragraph{Viewing disclosed vulnerabilities}
Perhaps because they are integral to learning (an important motivator), \hunters (\I{11}) appreciate being able to easily \emph{view a program's disclosed vulnerabilities} through the platforms (e.g., \emph{Hacktivity} on  Hackerone) (\FL{15}, \w{0.137}). Interviews suggest platforms are the primary source for accessing disclosed reports. A few (\I{5}) mentioned finding independent write-ups (e.g., blog posts); however, more \hunters mentioned disclosures made through \bbplatforms (\I{11}).

\subsection{Legal safe harbor}
\label{sec:legal_safe_harbor}

\textit{Legal safe harbor} is a commitment from companies not to 
legally pursue \hunters who follow the \change{published}{}rules, allowing \hunters to attack real world targets\change{ without repercussions}. \change{}{This legal aspect of \bbs was mentioned by \firststudyps under multiple \hlqs: choosing programs, challenges, and benefits. However, it is frequently overlooked in prior work.}

\change{\paragraph{Protection against legal threats}}{}
Though legal safe harbor was not listed as a benefit by many in the \firststudy (\FL{4}), it was ranked as relatively important (\change{\mean{5.96}}{\w0.118}). 
Similarly, safe harbor was not listed by many as a factor in choosing 
new programs (\FL{4}), but was again ranked as fairly important \change{\mean{5.71}}{(\w{0.098})}. %

Most \interviewps (\I{16}) said that \bbs without safe harbors 
are risky; six said they would not consider programs without them. 
\change{}{This might partly explain why legal safe harbor was not necessarily top of mind during the \firststudy; it may not be very visible, but its absence is noticeable.}

Some were less strict, saying they only avoid programs that 
have previously sued \hunters (\I{4}). ``I just need to hack on any program 
that they actually don't sue people. It makes my life easier.''\change{, I don't need 
to use a VPN [for anonymity].''}{} Eight recalled instances of legal 
action taken against \hunters, and two warned that companies without 
safe harbor policies would be likely to sue.

As another subtle benefit (\I{2}), safe harbor 
signals that \hunters are not malicious, 
as often portrayed~\cite{clark2021missourihacking}. 
One interviewee said, \change{``The value of it is more to remind the company, 
`These people are here to help you, not for you to get them in trouble.'}{``}So by this [safe harbor] existing, \ldots everybody acknowledges, `These hackers are good-guy hackers.' ''

Not all \hunters found legal safe harbor to be essential (\I{5}), 
perhaps related to \emph{legal threats} being seen as the second least important challenge (\FL{2}, \w{0.028}). \Interviewps explained why: it's already the law where they live (\I{1}); all 
HackerOne programs should already include safe harbors, so double-checking is 
unnecessary (\I{1}); and legal threats \change{related to \bbs}{}are rare (\I{1}). %
Four said it was the \hunter's responsibility to keep their activities legal. 
One added, ``I know my rights as a researcher. Even if you don't have a legal safe harbor, they have to prove some sort of damage.''%

\definecolor{Gray}{gray}{0.95}

\change{\Large{6 Clustering Bug Hunters{RQ3}}}{}
\section{Discussion}
\label{sec:discussion}
\change{}{When the ecosystem functions well, \bbs have the potential to improve the security posture of organizations (commercial or governmental) at a low cost~\cite{sridhar2021hacking, walshe2020empirical}, while also providing \hunters numerous benefits. However, \bbs have low adoption rates, and only some organizations that run \bbs consistently attract quality bug reports~\cite{maillart2017given, sridhar2021hacking}. On the \hunters' side, while the reported pool of workers is large, only a few receive the full breadth of promised benefits~\cite{ellis2022bounty}. Essentially, not only are \bbs underutilized, but there exist inequalities in how the current ecosystem's benefits are distributed among organizations and \hunters.}

\change{}{In this section, we list recommendations for \bbprograms, platforms, and policymakers to alleviate the important challenges we have identified, as well as support and expand the benefits \hunters appreciate the most. We hope this can bolster \bb adoption, helping the security of more organizations, and ensuring more \hunters receive the full benefits of this marketplace.}

\paragraph{What \bbprograms can do} The two highest-ranked factors that programs have direct control over are scope and rewards, but both require additional resources to increase.
Rewards are directly tied to finances, while an increased scope might increase staffing requirements; we suggest future work explore the nuances of increasing scope. %
A larger scope, when feasible, is also likely to alleviate some other concerns \hunters have, such as duplicates, saturation, and the range of technologies used.
Closely related to the extent of the scope, the clarity of the scope was also seen as an important issue. Making sure that the scope is clear and up-to-date should be relatively low-cost and could reduce \hunters' frustration
with unexpected out-of-scope and invalid responses.  %

\change{\Bbprograms have much room to improve in all aspects.}{The most significant challenges hunters reported were all related to communication.}
To address responsiveness, the foremost challenge, \change{}{increased staffing is likely to be the most effective solution~\cite{ellis2022bounty}. If additional staffing is not possible, a cheaper option is to provide frequent and transparent updates on the status of a bug report to reduce uncertainty (\autoref{subsec:disputes}).}

Unexpected responses could also be reduced by improving communications overall. \Bbprograms could start by making scopes and bounty tables as clear as possible, perhaps including examples with payouts per (publicly disclosed) bug and by drawing attention to common out-of-scope submissions and pitfalls. Programs could also train their 
managers to improve communications with \hunters and avoid common pitfalls 
in interpreting vulnerability reports.

\paragraph{What \bbplatforms can do} 
Poor platform support (\ie mediation) was the most significant platform issue, and dissatisfaction with platform or program responses (leading to disputes and then potentially to mediation) is the second most important challenge overall. Several participants complained that mediators are biased against \hunters. Platforms could address this by more clearly communicating their business models---they need \hunters as much as \bbprograms to function---and \change{making their decision process transparent}{increasing transparency by creating periodic reports on the outcomes of mediation}. 

More generally, platforms could also adopt several of our above suggestions 
for improving communications issues, including adding more staff 
where possible to reduce delays, and adding guidance for submitting reports. Platforms could also offer training in how to communicate with \hunters and interpret vulnerability reports---based on prior examples on the platform--- to participating programs. %

\change{}{We also suggest platforms explore how to reduce the uncertainty \hunters face.} Platforms could \change{}{evaluate} the utility of implementing insurance policies to ensure \hunters are paid even if \bbprograms are unwilling to accept the results of mediation; or, more aggressively, attempt to reduce the authority of \bbprograms to issue final judgements. 
\change{}{For instance, platforms could consider managing more of the triaging process and even deciding on final payments.} %

Learning was seen as the second most important benefit of \bbs \change{by a margin}{ and perhaps the primary way productive \hunters are added to the workforce}\change{; however,}{. Currently, only a few \hunters enjoy the full benefits of the \bb ecosystem; however, increasing \hunters' skill levels could reduce this gap, while also leading to more bugs being found overall, and therefore potentially better software security in general.} Unfortunately, learning resources provided by \bbplatforms were the third least important feature, indicating room for improvement. We suggest better integrating learning material with previously disclosed bugs (the third most useful platform feature), in order to make this material more useful and relevant. %

\change{}{Beyond improving less popular aspects, \bbplatforms should not neglect their most popular features: providing a wide range of payment mechanisms (e.g., for international \hunters), standardizing interfaces between programs, and enabling \hunters to review (and read others' reviews of) \bbprograms.}

\Bbplatforms could also help distribute \hunters' attention among \bbprograms by promoting low-attention but well maintained or societally important programs, \change{although it's unclear whether they currently have incentives to do so.}{though this may need to be incentivized by an outside party (as we discuss next).}

\paragraph{\change{}{What legislative bodies and policymakers can do}}
\change{}{
Bug-bounty programs offer important security benefits that could be 
useful to many companies. However, our results suggest \hunters 
typically focus on the programs with the most resources (\eg monetary 
rewards, large scopes). Large companies (e.g., Google), 
with committed \bbprograms, are well positioned to take 
advantage in ways smaller companies and programs may not be able to, 
even when these companies are in critical sectors with important security 
needs. Exacerbating the issue, smaller companies are at higher risk for suffering significant loss to cybercrime~\cite{hauman2021smesec, sridhar2021hacking}.}

\change{}{Several government agencies currently recommend companies adopt \bbs~\cite{nist2020bugbounties,cisa2020bugbounties}. While this is a good start, without additional support, our results suggest it has the effect of entrenching security inequality.
Government agencies could also provide funding to help companies with lower security budgets and staffing afford to run committed \bbprograms: increasing scope, better staffing, and higher bounties. Grants (e.g.~\cite{EUgrants}) could be tied to priorities for improving the \bb ecosystem, such as publicizing reports (enabling \hunters to assess programs), attracting attention to less popular but security-critical industries (e.g., healthcare, finance~\cite{sridhar2021hacking}), or improving educational resources~\cite{votipka2021hacked}.}

\change{}{Governments have additional roles to play in improving conditions for bug hunters. Legal scholars have proposed legislation to implement legal safe harbors for any good-faith security researcher, including \hunters~\cite{etcovitch2018coming}, and judicial policy-setters have taken steps in this direction~\cite{doj2022goodfaith}. Our work provides evidence that this is a worthwhile effort, as \hunters value a legal safe harbor and use it as a discriminator between programs (\autoref{sec:legal_safe_harbor}). A generalized safe harbor could therefore incentivize \hunter participation for companies that do not currently offer that benefit. %
Further, labor regulators could consider how best to protect \hunters from uncertainty and even abuse associated with gig-work (\autoref{sec:freelance}, \cite{ellis2022bounty}), by setting appropriate standards for communication and even payment, which would help to retain more \hunters in the ecosystem.}

\change{\large{7 Conclusion}}{}

\section*{Acknowledgments}
We thank our participants, without whom this study would not be possible; and our reviewers, who helped shape the analysis and framing of the paper. We also thank Jack Cable, Julien Ahrens, and Nathan Reitinger for their assistance and insights.

This material is based upon work sponsored by the National Science Foundation under Grants No.\ CNS-1850510 and CNS-1801545.
Any opinions, findings, and conclusions or recommendations expressed in this material are those of the authors and do not necessarily reflect the views of the National Science Foundation. This research was also supported by a gift from Google.

\bibliographystyle{plain}
\bibliography{main}

\appendix

\clearpage

\ifIsArxiv

\section{Additional Factors}
\label{app:additional_factors}

In the following, we discuss additional factors that \hunters consider when participating in \bbs. 

\subsection{Other benefits}

\paragraph{Enjoyment}
A highly ranked benefit for \hunters was simply the enjoyment of participating in \bbs (\FL{20}, {\w0.140}). Specifically, \interviewps mentioned intellectual stimulation (\I{3}): ``It's fun to break something. \ldots %
It's like a challenge. It's a puzzle.'' Others mentioned the thrill of 
beating the development team (\I{1}), bug hunting as relaxation (\I{1}), and even 
the thrill of uncertainty (\I{1}). ``It's like gambling, is an analogy I'd often use. You don't really know if you get the next payout. I guess at the same time, that's what makes it fun.'' One participant specifically noted that \bbs were only enjoyable as a hobby and would lose their appeal as a full-time job.

\paragraph{Altruism}
A minority of participants in the \firststudy said they participate in \bbs for \textit{altruistic} reasons (\FL{5}). But \secondstudyps did not rank altruism as a top benefit (\w{0.062}). Four \hunters saw \bbs as an outlet to make the internet secure for other users. As one said, ``Altruism and hacking for good is, I think, one of the main roles that bug bounty platforms and ethical hacking even exist.'' One saw sharing knowledge to help other \hunters to be altruistic. Four said they would report a bug regardless of whether they would be paid, and one said they would help a company that could not afford to pay: ``I don't mind doing pro bono work, but that's only really when people absolutely can't afford me.'' 
However, the same participant noted that such companies usually ``don't have the 
maturity to [take advantage of] the advice I give them, which is kind of a shame.''
Other \Interviewps (\I{4}) contrasted their beneficent work with harms 
inflicted by criminal hackers.

\subsection{Other challenges}

\paragraph{Language barriers}
One \firststudyp identified language barriers between \hunters and \managers as another communication problem. None of our \interviewps mentioned this; however, we note that
all \interviewps were by necessity sufficiently fluent in English to be interviewed.
Our \interviewps might be a biased sample as we were able to interview all of them in English.

\change{
\paragraph{Altruism}
A minority of participants in the \firststudy said they participate in \bbs for \textit{altruistic} reasons (\FL{5}). \secondstudyps did not rank altruism as a top benefit (\mean{5.54}) Four \hunters saw \bbs as an outlet to make the internet secure for other users. As one said, ``Altruism and hacking for good is, I think, one of the main roles that bug bounty platforms and ethical hacking even exist.'' One saw sharing knowledge to help other \hunters to be altruistic. Four said they would report a bug regardless of whether they would be paid, and one said they would help a company that could not afford to pay: ``I don't mind doing pro bono work, but that's only really when people absolutely can't afford me.'' 
However, the same participant noted that such companies usually ``don't have the 
maturity to [take advantage of] the advice I give them, which is kind of a shame.''
Other \Interviewps (\I{4}) contrasted their beneficent work with harms 
inflicted by criminal hackers. }{}

\change{\paragraph{Platform prepared educational material}
\Hunters also obtained new knowledge through  educational materials provided by \bbplatforms ($n$=15).  This included videos like those in Bugcrowd University\footnote{\url{https://www.bugcrowd.com/hackers/bugcrowd-university/}} ($n$=8), hacking exercises like the Hacker101 CTF~\footnote{\url{https://ctf.hacker101.com/}} ($n$=4), platform-prepared blog posts ($n$=3), platform-sponsored events ($n$=1), and discounts on hacking tools through the platforms ($n$=1). 
However, there was some debate regarding whether \bbplatforms produced material appropriate for all skill levels. For example, one participant explained that while they were valuable when starting, they mostly provide a jumping off point with little depth, saying ``And I also watched the Hacker 101 videos which got me introduced in that more specifically. But I think that doesn't really not goes too deeply into the theoretical aspects \ldots''
On the other hand, one \interviewp thought \bbplatforms produced educational content for \hunters in all career stages.}{}

\paragraph{Too much menial labor} 
In the \firststudy, twelve participants noted their annoyance with 
repetitive, menial tasks commonly required by \managers. 
\Interviewps described two key sources of frustration: company network 
architecture and required testing accounts or credentials.

With respect to network architecture, \interviewps complained about 
needing to bypass content-delivery networks (\I{2}), firewalls (\I{2}), 
or CAPTCHAs (\I{1}). \Hunters found it frustrating that \bbprograms 
conflate bypassing these (often hardened) tools with finding bugs at 
the target itself: ``I'm happy to do a web application firewall 
assessment, but you're asking me to do a website assessment. \ldots~If they're using CloudFlare [for firewalling], am I bypassing CloudFlare? 
I'm not really going to get the right kind of rewards from Barry's 
cake shop that I've just bypassed a CloudFlare controller and 
I've got a \$500 XSS. When actually the CloudFlare bypass might be worth more to CloudFlare.''

Requiring test accounts---and specifically VPNs---raises privacy concerns (\I{2}): 
``I do not like connecting to VPNs and letting them monitor my traffic. \ldots I'm okay with going through a proxy \ldots to bypass the web application firewall. That's still reasonable, but tunneling everything through a VPN, just kind of cringe on that.''
Notably, previous research has shown VPNs to be deceptive and detrimental to privacy~\cite{ikram2016analysis}.
Others (\I{2}) noted that setting up test accounts can be a manual and finicky 
process. For example, one \hunter recalled difficulty obtaining test accounts 
from a social media company: ``I found some bug on [website], and their automated firewall kept on blocking me \ldots so I stopped doing research at their website.''

Finally, two \hunters made note of annoyances stemming from their geographic locations; one noted that some websites do not allow international users to register even though their \bbprograms are international. Interestingly, another hacker was complaining about region-specific formatting, ``That's the part that makes me start to test more [local] programs. For example, there was a scope that I never tested in America, because I needed some address and I use it that generator websites. And for some reason, some information wasn't working.''

In contrast, two \hunters found menial tasks useful. 
One said the configuration process (including setting up 
test accounts) allows the \hunter to become familiar with 
the system, and another suggested these menial tasks 
themselves as a viable attack target, essentially widening 
the scope. A third argued that test accounts are simply 
necessary: ``you can hack a lot of different websites. But if you don't exploit it in the correct way, you can easily crush the website.''

\subsection{Intrinsic company factors}
Some factors that \hunters consider are intrinsic to the companies that host \bbprograms. 
Although none were rated as a top discriminator, we identify three such factors in the \firststudy, the business domain of the company (\FL{2}, \w{0.021}),
what country the company is based out of (\FL{1}, \w{0.006}), and how popular the company is (\FL{15}, \w{0.047}).

\paragraph{Business domain}
The business domain (e-commerce, crypto-currencies, government, etc.) of the company hosting a \bbprogram was listed as a factor considered when choosing what program to work on by two \firststudyps.

\Interviewps generally had practical or emotional considerations \cite{davis1985technology} when interpreting the business domain of a \bbprogram. On the practical side, an \interviewp preferred to hack finance businesses assuming they had more to lose and increasing bug impact. Another avoided government defense, law enforcement, and intelligence agencies; they were under the impression that these agencies would investigate any \hunter that worked on their programs, ``I tend to avoid government targets like the CIA and FBI\ldots{}I don't want those interrogations at the border, so I tend to avoid them.'' The fourth \interviewp to list practical reasons preferred not to hack blockchain companies. This participant had several reasons for not doing so, ``The technology is not interesting to me\ldots{}Because of how young they are, they will not be very mature in terms of how they respond to [communications] \ldots{} they'll pay in cryptocurrency\ldots{} tax costs of that, it's just not worth it at all.''

Listing emotional reasons, one hacker noted ethical reasons for not working for the aviation industry, ``I can tell you that I refuse to work on airlines, for example. It's for ethical reasons. \ldots~I firmly believe that we need to reduce our environmental impact. I think people should be flying less.'' Another hacker avoided companies that carried social stigma, ``Yeah. I think there's only a few programs actually that I think would have a negative stigma, at least from my perspective. For example, like PornHub or Roblox.''

\paragraph{Country}
One participant in the \firststudy indicated that they cared about the company's home country (e.g., Uber is a U.S. based company). Of the \interviewps, one noted that working with companies in their own country provided competitive advantages,
``I think that is because there is less people searching, not because the place, but because it's more difficult \ldots{} 
for someone outside of Brazil to make an account and complete the documents and other things that will be necessary to get in some places.'' Another noted they preferred to hack programs from their own country (India) because (1) they had more vulnerabilities and (2) the company was based out of their own country. Overall, it seems that some \hunters prefer \bbprograms from certain countries due to perceived personal advantages gained often through physical proximity.

The opposing view was that \bbs is remote work, therefore, it does not matter where you are in world.

\paragraph{Company familiarity}
Familiarity with a company or the products of a company was one of the more often mentioned factors in choosing \bbprograms to work on (\FL{15}).

Of the \interviewps that talked about this factor (\I{9}), the majority said they cared about it because familiarity helped them with reconnaissance of the assets a \bbprogram included (\I{8}). A \hunter explained, ``I like to use the site or the app as its meant to be used \ldots Then it's going to give you ideas of how this could be abused or whether something isn't working as intended. \ldots If you're already familiar with it, then that's always a step that you can skip.''
Two \hunters had a more specific strategy, they kept track of acquisitions of companies. They argued that new acquisitions eventually get added into the scope of \bbprograms. By tracking new acquisitions, \hunters can get a headstart investigating new targets before they are publicly added to the scope. One \hunter explained this in the context of the Google VRP~\footnote{\url{https://bughunters.google.com/}}, ``Google pays bounties on their acquisitions, as long as they've been in acquisition for six months. So you can go and do a ton of recon and even find vulnerabilities per se or what you might suspect is a vulnerability. And then if they've been around for six months, you can report those.'' Finally, a \hunter mentioned they preferred hacking---and therefore eventually securing---products they personally use.

\fi %

\ifIsArxiv

\section{\Firststudy Survey}
\label{app:firststudysurvey}

[Participants are directed from recruitment material that explains the study.]

\noindent[Show consent form.]

\noindent[Do not proceed if consent is not given.]\\

\noindent[Main measurement questions:]

\begin{itemize}
    \item What are all the factors that you consider when choosing which bug-bounty programs to participate? Please list every factor you have considered even if it is not the case that you think about it every time you pick a bug-bounty program. Please keep trying to recall factors if you think there are more that you might be able to remember.
    
    Please place each factor on a new line.
    
    [Free-text response]
    
    \item What are all the issues that make you stop working on a particular bug-bounty program? This could be anything related to your relationship with the program organizers, the system that you are investigating, or some other external factor. Again, please keep trying to recall issues if you think there are more that you might be able to remember.

    Please place each issue on a new line.
    
    [Free-text response]
    
    \item What are all the benefits of working on bug-bounty programs for you? Please list all the reasons why you participate in bug bounty programs. Again, please keep trying to recall reasons for participation if you think there are more that you might be able to remember.

    Please place each reason on a new line.
    
    [Free-text response]
    
    \item What are all the challenges that you face working on bug-bounty programs? What factors make working on a bug-bounty program difficult. These can be related to the program’s organization, the system you’re investigating itself, or due to other factors. Please keep trying to recall challenges if you think there are more that you might be able to remember. What do you do to overcome these challenges?

    Please place each challenge on a new line.
    
    [Free-text response]
    
    \item What are the most useful features of the bug bounty platforms you use? Please keep trying to recall features if you think there are more that you might be able to remember.

    Please place each feature on a new line
    
    [Free-text response]

    \item What changes would you like platforms and programs to implement?
    
    [Free-text response]

\end{itemize}

[Demographics and experience.]
\begin{itemize}
    \item On a scale from 1-5, how would you assess your vulnerability discovery skill (1 being a beginner and 5 being an expert)?
    
    [An integer slider between 1-5.]
    
    \item Please select the range which most closely matches the number of software vulnerabilities have you discovered?
    \begin{itemize}
        \item 0-3
        \item 4-6
        \item 7-10
        \item 11-25
        \item 26-50
        \item 51-100
        \item 101-500
        \item > 500
    \end{itemize}
    \item How many total years of experience do you have with vulnerability discovery?
    
    [Free-text response]
    \item Please select the range that most closely matches the amount of time you typically spend performing software vulnerability discovery tasks per week.
    \begin{itemize}
        \item < 5 hours
        \item 5-10 hours
        \item 10-20 hours
        \item 20-30 hours
        \item 30-40 hours
        \item > 40 hours
    \end{itemize}
    
    \item Please specify the range that closely matches the amount of time you typically spend on non-vulnerability discovery, technical task per week (e.g. software or hardware programming, system administration, network analysis etc)?
    \begin{itemize}
        \item < 5 hours
        \item 5-10 hours
        \item 10-20 hours
        \item 20-30 hours
        \item 30-40 hours
        \item > 40 hours
    \end{itemize}
    
    \item Please specify the gender with which you most closely identify.
    \begin{itemize}
        \item Male
        \item Female
        \item Other
        \item Prefer not to answer
    \end{itemize}
    
    \item Please specify your age.
    \begin{itemize}
        \item 18-29
        \item 30-39
        \item 40-49
        \item 50-59
        \item 60-69
        \item > 70
    \end{itemize}
    
    \item Please specify your ethnicity
    \begin{itemize}
        \item White
        \item Hispanic or Latino
        \item Black or African American
        \item American Indian or Alaska Native
        \item Asian, Native Hawaiian, or Pacific Islander
        \item Other
    \end{itemize}
    
    \item Please specify which country/state/province you live in.
    
    [Free-text response]
    
    \item Please specify the highest degree or level of school you have completed
    \begin{itemize}
        \item Some high school credit, no diploma or equivalent
        \item High school graduate, diploma or the equivalent (for example: GED)
        \item Some college credit, no degree
        \item Trade/technical/vocational training
        \item Associate degree
        \item Bachelor's degree
        \item Master's degree
        \item Professional degree
        \item Doctorate degree
    \end{itemize}
    
    \item If you are currently a student or have completed a college degree, please specify your field(s) of study (e.g. Biology, Computer Science, etc).
    
    [Free-text response]
    
    \item Please select the response option that best describes your current employment status.
    \begin{itemize}
        \item Working for payment or profit
        \item Unemployed
        \item Looking after home/family
        \item A student
        \item Retired
        \item Unable to work due to permanent sickness or disability
        \item Other [free-text response]
    \end{itemize}
    
    \item If you are currently working for payment, please specify your current job title.
    
    [Free-text response]
    
    \item Please specify the range which most closely matches your total, pre-tax, household income in 2018.
    \begin{itemize}
        \item < \$29,999
        \item \$30,000 - \$49,999
        \item \$50,000 - \$74,999
        \item \$75,000 - \$99,999
        \item \$100,000 - \$124,999
        \item \$125,000 - \$149,999
        \item \$150,000 - \$199,999
        \item > \$200,000
    \end{itemize}
    
    \item Please specify the range which most closely matches your total, pre-tax, household income specifically from vulnerability discovery and software testing in 2018.
    \begin{itemize}
        \item < \$29,999
        \item \$30,000 - \$49,999
        \item \$50,000 - \$74,999
        \item \$75,000 - \$99,999
        \item \$100,000 - \$124,999
        \item \$125,000 - \$149,999
        \item \$150,000 - \$199,999
        \item > \$200,000
    \end{itemize}
\end{itemize}

[Contact for future studies consent]

\begin{itemize}
    \item Please indicate whether you would be ok with us contacting you regarding future studies.
    \begin{itemize}
        \item I agree to be contacted regarding future studies
        \item I do not agree to be contacted regarding future studies
    \end{itemize}
\end{itemize}

We thank you for your time spent taking this survey.  Your response has been recorded.

If you would like to learn more about our research, please check out our website: [redacted].  At the conclusion of our study, we will provide a link to our published results on our website.

\fi

\ifIsArxiv

\section{\Secondstudy Survey}
\label{app:secondstudysurvey}

[Participants are directed form recruitment material that explains the study.]

\noindent[Show consent form.]

\noindent[Do not proceed if consent is not given.]\\

\noindent This survey consists of three parts:
 
\begin{enumerate}
    \item Questions about factors surrounding bug bounty hunting and your opinions of them
    \item Questions about your experience with bug bounty hunting
    \item Demographics questions.
\end{enumerate}

In the next section, you will be asked to rank the importance of certain factors surrounding bug bounty programs. There are 4 questions in this section.

\begin{itemize}
    \item How important are the following factors to you when choosing in which bug-bounty programs to participate?

    [List all ``Choosing a program'' factors and definitions in a Likert matrix as it appears in \autoref{app:full_list} in randomized order.]
    
    [Scale: Extremely important - Very important - Moderately important - Neutral - Slightly important - Low importance - Not at all important]
    
    \item How significant are the following challenges of working on bug-bounty programs to you?

    [List all ``Challenges of bug hunting'' and definitions in a Likert matrix as it appears in \autoref{app:full_list} in randomized order.]
    
    [Scale: Extremely challenging - Very challenging - Moderately challenging - Neutral - Slightly challenging - Low challenge - Not at all challenging]
    
    \item How important are the following benefits of working on bug-bounty programs to you?
    
    [List all ``Benefits of bug hunting'' and definitions in a Likert matrix as it appears in \autoref{app:full_list} in randomized order.]
    
    [Scale: Extremely important - Very important - Moderately important - Neutral - Slightly important - Low importance - Not at all important]

    \item How useful are the following features of bug-bounty platforms (e.g. HackerOne, Bugcrowd) to you?
    
    [List all ``Useful platform features'' and definitions in a Likert matrix as it appears in \autoref{app:full_list} in randomized order.]
    
    [Scale: Extremely useful - Moderately useful - Slightly useful - Neither useful nor useless - Slightly useless - Moderately usefless - Extremely useless]

\end{itemize}

[Skills and experience]

In the next section, you will be asked questions about your bug bounty hunting experience.
\begin{itemize}
    \item How did you first get involved with bug bounty programs?
    
    [free-text response]
    
    \item How would you assess your skill level as a bug bounty hunter on the following scale?
    \begin{itemize}
        \item 1 - Fundamental Awareness (basic knowledge)
        \item 2 - Novice (limited experience)
        \item 3 - Intermediate (practical application)
        \item 4 - Advanced (applied theory)
        \item 5 - Expert (recognized authority)
    \end{itemize}
    
    \item How many software vulnerabilities have you discovered for which you received bug bounty rewards (including non-monetary rewards)?
    
    [free-text numeric response]
    
    \item How many years of experience do you have working with bug bounty programs?

    [free-text numeric response]

    \item Which bug bounty platforms (i.e., the companies that host many bug bounty programs) have you ever worked on?
    
    [free-text response]

    \item Which bug bounty programs have you worked on recently?
    
    [free-text response]
    
    \item Which range matches most closely the amount of time that you typically spend per week working on bug bounty programs?
    \begin{itemize}
        \item < 5 hours
        \item 5-10 hours
        \item 10-20 hours
        \item 20-30 hours
        \item 30-40 hours
        \item > 40 hours
    \end{itemize}
    
    \item Which range matches most closely the amount of time that you typically spend per week on technical tasks that are not related to bug bounty (software or hardware programming, system administration etc.)?
    \begin{itemize}
        \item < 5 hours
        \item 5-10 hours
        \item 10-20 hours
        \item 20-30 hours
        \item 30-40 hours
        \item > 40 hours
    \end{itemize}   

\end{itemize}

[Demographics questions]

In the next section, you will be asked some demographics questions.

\begin{itemize}
    \item With which gender do you most closely identify?
    \begin{itemize}
        \item Male
        \item Female
        \item Other [free-text]
        \item Prefer not to answer
    \end{itemize}

    \item How old are you?
    \begin{itemize}
        \item 18-29
        \item 30-39
        \item 40-49
        \item 50-59
        \item 60-69
        \item > 70
        \item Prefer not to answer
    \end{itemize}
    
    \item In which country do you currently reside?
    \begin{itemize}
        \item USA
        \item India
        \item Russia
        \item Germany
        \item Canada
        \item United Kingdom
        \item Sweden
        \item Netherlands
        \item China
        \item Australia
        \item Other [free-text]
        \item Prefer not to answer
    \end{itemize}
    
    \item What is the highest degree or level of school you have completed?
        \begin{itemize}
        \item Some high school credit, no diploma or equivalent
        \item High school graduate, diploma or the equivalent (for example: GED)
        \item Some college credit, no degree
        \item Trade/technical/vocational training
        \item Associate degree
        \item Bachelor's degree
        \item Master's degree
        \item Professional degree
        \item Doctorate degree
        \item Other [free-text]
        \item Prefer not to answer
    \end{itemize}
    
    \item If you are currently a student or have completed a college degree, what is / was your field(s) of study (e.g. Biology, Computer Science)?
    
    [Free-text response]
    
    \item Which option describes your current employment status best?
        \begin{itemize}
        \item Working for payment or profit
        \item Unemployed
        \item Looking after home/family
        \item A student
        \item Retired
        \item Unable to work due to permanent sickness or disability
        \item Other [free-text response]
        \item Prefer not to answer
    \end{itemize}
    
    \item If you are currently employed, what is your current job title?
    
    [Free-text response]
    
    \item Which range matches most closely your total, pre-tax household income in 2019?
    \begin{itemize}
        \item < \$29,999
        \item \$30,000 - \$49,999
        \item \$50,000 - \$74,999
        \item \$75,000 - \$99,999
        \item \$100,000 - \$124,999
        \item \$125,000 - \$149,999
        \item \$150,000 - \$199,999
        \item > \$200,000
        \item Prefer not to answer
    \end{itemize}
    
    \item Which range matches most closely your total, pre-tax income from bug bounties in 2019?
    \begin{itemize}
        \item < \$29,999
        \item \$30,000 - \$49,999
        \item \$50,000 - \$74,999
        \item \$75,000 - \$99,999
        \item \$100,000 - \$124,999
        \item \$125,000 - \$149,999
        \item \$150,000 - \$199,999
        \item > \$200,000
        \item Prefer not to answer
    \end{itemize}
\end{itemize}

[Contact for future studies consent]

\begin{itemize}
    \item Would you be OK with us contacting you regarding future studies?
    \begin{itemize}
        \item I agree to be contacted regarding future studies
        \item I do not agree to be contacted regarding future studies
    \end{itemize}
\end{itemize}

Thanks for your response! We will reach out to you for the interview part of the study based on your response. Meanwhile, please refer to our website [redacted] if you have any questions.

\fi

\definecolor{Gray}{gray}{0.95}

\begin{table*}[hbpt!] %
\section{Complete List of Factors}
\label{app:full_list}

  \centering
  \resizebox{\textwidth}{!}{
  \footnotesize
  \begin{tabular}{ p{0.025\textwidth} r p{0.65\textwidth}  p{0.01\textwidth}  p{0.03\textwidth} p{0.03\textwidth}  }
  \toprule
  \small{\textbf{Question}} & \small{\textbf{Code Name}} & \small{\textbf{Description} } & \small{\textbf{FL}} & \small{\textbf{$\mu_r$}} & \small{\textbf{Worth ($\pi$)}}  \\
  \midrule
  \multirow{20}{*}{\small{\textbf{\rotatebox[origin=c]{90}{Choosing a program }}}} &  
  
        \textbf{Scope:} & Number of domains or assets that are included in the program. & 28 & 6.10 & 0.156 \\
        &  \textbf{Reward:} & Expected monetary or non-monetary rewards (e.g., SWAG, hardware, subscription). & 36 & 5.91 & 0.120 \\
        & \textbf{Bounty table:} & Reward rules and ranges set by the managers (e.g., \$50 for low criticality bugs, but \$5000 for high criticality bugs). & 16 & 5.87 & 0.117 \\
        &  \textbf{Technology familiarity:} & Familiarity with the technology of the assets (e.g., familiarity with web or iOS). & 22 & 5.86 & 0.113 \\
        & \textbf{Legal safe harbor:} &  Language of program includes a commitment to not pursue legal actions after hackers who follow the rules and/or explicitly authorizes testing conducted in accordance with the rules. & 4 & 5.71 & 0.098 \\
        &  \textbf{Program repute:} & Program’s reputation in the community for being pleasant to work with (i.e., what other hackers say about the program). & 15 & 5.68 & 0.086 \\
        & \textbf{Learning opportunity:} & Lack of familiarity with the technology of the assets (e.g., interest in learning crypto or Android). & 1 & 5.35 & 0.064\\
        &  \textbf{Private or public:} & Private programs (accessible only by invitation) vs. public programs (accessible by anyone). & 4 & 5.16 & 0.048\\
        & \textbf{Company familiarity:} & Company behind the program is widely known, or you or your peers use its products or services (e.g., working on Uber’s program because you like or use their services). & 15 & 5.04 & 0.047 \\
        &  \textbf{Saturation:} & Number of reports received or number of hackers working on the program. & 8 & 5.17 & 0.047 \\
        
        & \textbf{Career opportunities:} & Future career opportunities with the company behind the program. & 3 & 4.49 & 0.027 \\
      
        &  \textbf{Public disclosure:} & Public vulnerability disclosure is generally allowed following the resolution of the issue, permissive NDAs. & 6 & 4.63 & 0.027 \\ 
        
        & \textbf{Age:} & For how long the program has been running.  & 6 & 4.44 & 0.023 \\

        &  \textbf{Business domain:} & Business domain of the company behind the program (e.g., social media, insurance, medical). & 2 & 4.47 & 0.021 \\

        & \textbf{Country:} & Where the company behind the program is located. & 1 & 3.34 & 0.006\\

\midrule

    \multirow{19}{*}{\small{\textbf{\rotatebox[origin=c]{90}{Challenges of bug hunting}}}} &
    
      \textbf{Poor responsiveness:}& Lack of responses or slow responses from program managers. & 30 & 5.39 & 0.130 \\  
    &  \textbf{Dissatisfaction with responses:}& Rewards are lower than promised by rules (e.g., downgraded severity, impact, disagreements with duplicates). & 26 & 5.36 & 0.120 \\
    
    & \textbf{Unclear scope:} & Program scope is not defined clearly. & 3 & 4.99 & 0.082 \\

    &  \textbf{Poor platform support}& Dissatisfaction with how platforms handle issues, such as mediating between hackers and programs. & 1 &  5.03 & 0.079   \\

    &  \textbf{Duplicates:}& Too many reports marked as duplicates. & 4 & 5.02 & 0.078 \\

    &  \textbf{Assets outside expertise} & Assets are outside area of expertise, lacking certain required skills.&  11 &  4.87 & 0.068 \\
    
    &   \textbf{Secure assets:} & Finding bugs is too difficult. & 5 & 4.78 & 0.064 \\
    
    &   \textbf{Stress and uncertainty:} & Fear of burning out, social isolation during work, irregular income, etc. & 5 & 4.8 & 0.062 \\

    &   \textbf{Too much labor work:} & Menial tasks (e.g., CAPTCHA, waiting for timeouts, obfuscation, setting up test accounts). & 12 & 4.62 & 0.050 \\

    &  \textbf{Boredom:}& Bored of working on the program or a more interesting program launches. & 8 & 4.62 & 0.050 \\
    
    & \textbf{Unrepresentative reputation system:} & Hackers' reputation points do not reflect real experience and are not transferable between platforms.  & 1 & 4.59 & 0.047 \\

    &  \textbf{Difficulty working with managers:} & Bug-bounty program managers are difficult to work with (e.g., disrespectful, requiring extra work). & 23 &  4.48 & 0.044\\

    & \textbf{Not enough time:}& Not having enough time for participating in bug bounties. & 2 & 4.45 & 0.043 \\

    &  \textbf{Limited vulnerability disclosure:}& Restrictive vulnerability disclosure policies and NDAs that may prevent you from publishing your work following the resolution/mitigation of the issue. & 2 & 4.49 & 0.041 \\
    
    & \textbf{Legal threats:}& Fear of threats of legal implication (civil or criminal).  & 2 & 3.96 & 0.028 \\
    
    &   \textbf{Lacking communication or language skills:} & Communication difficulties because you feel that you lack language skills, experience anxiety in communication, etc. & 2 & 3.45 & 0.014 \\

\midrule

     \multirow{10}{*}{\small{\textbf{\rotatebox[origin=c]{90}{Benefits of bug hunting}}}}
     &  \textbf{Monetary rewards:} & Monetary compensation. & 42 & 6.31 & 0.191 \\
     &   \textbf{Learning:} & Learning or improving skills. & 32 & 6.18 & 0.170 \\
     &  \textbf{Enjoyment:}& Enjoyment or challenge of white-hat hacking. & 20 & 6.08 & 0.140 \\
     &   \textbf{Legal safe harbour:}& Hacking without the threat of legal actions if they obey the rules. & 4 & 5.96 & 0.118\\

     &  \textbf{Flexibility:}& Work schedule and place flexibility (compared to traditional employment). & 16 & 5.85 & 0.095 \\
     &  \textbf{Career:}& Building relations and reputation with companies for employment and other work opportunities. & 11 & 5.71 & 0.091 \\
     &  \textbf{Community:}& Bug bounty creates a community of hackers. & 3 & 5.52 & 0.071 \\
     &   \textbf{Altruism:}& Improving cybersecurity for the sake of helping others, hacking to make the internet safer for everyone. & 5 & 5.54 & 0.062 \\
    
     &  \textbf{Reputation:}& Earning platform reputation points, building a following, etc. & 14 & 5.36 & 0.048 \\
     &   \textbf{Non-monetary rewards:}& Non-monetary compensation (e.g., SWAG, hardware, subscriptions). & 4 & 4.35 & 0.013 \\

\midrule

    \multirow{15}{*}{\small{\textbf{\rotatebox[origin=c]{90}{Useful platform features}}}}

    &  \textbf{Ease of payment:} & Receiving payments in a standardized, hassle-free way. & 12 & 6.51 & 0.156 \\[-1.2em]
    &   \textbf{Ease of reporting:} & Easy to generate, submit, and track reports and their status. & 16 & 6.46 & 0.142 \\
    & \textbf{Viewing disclosed vulnerabilities:} & Platform provided interface for viewing bugs found by others. & 15 & 6.41 & 0.137 \\
    &    \textbf{Private program invitations:} & Access to private programs on the platform. & 5 & 6.28 & 0.107  \\
    &   \textbf{Program directory:} & Listing many programs in one place, with statistics, details, etc. (being able to view Uber, Paypal, etc. programs on one page with statistics).  & 17 & 6.02 & 0.068 \\
    &   \textbf{Standardized rules:} & Platform standardizing how scopes, rewards, criticality, etc. are defined. & 3 & 5.99 & 0.063 \\
    &   \textbf{Community:} & Platform making effort to create a community of hackers. & 11 & 5.77 & 0.057 \\
    &   \textbf{Platform rewards:} & E.g., platform SWAG, funded travel. & 1 & 5.89 & 0.054 \\
    &   \textbf{Mediation:} & Platform resolving disputes between hackers and programs. & 13 & 5.77  & 0.051 \\
    &  \textbf{Platform managed disclosure:} & Platform provided tools/mechanisms to publicly disclose resolved bugs. & 6 & 5.86 & 0.050 \\
    &   \textbf{Resources for learning:} & Platform providing free resources on how to hack (e.g., Bugcrowd University). & 2 & 5.59 & 0.047 \\
    &   \textbf{Reputation system:} & Platform managed reputation system for hackers. & 6 & 5.70 & 0.043 \\

    &  \textbf{Platform triage:} & Triaging managed by the platform (e.g., HackerOne triages your report instead of Uber). & 5 & 5.06 & 0.023 \\
    
    &   \textbf{None:} & There are no useful features that platforms provide. & 4 & - & - \\

 \bottomrule
 \\

 \end{tabular}
}
\caption{\change{}{Factors used in the \secondstudy, organized by  \hlqs. FL: count of participants who listed the factor in the \firststudy. Worth ($\pi$): the estimated relative importance of factors (see~\autoref{subsec:ratingmethods} for details). Likert averages ($\mu_r$) were included to show differences with LLBT-based analysis.
}}
\label{tab:refined_codes_full}
\end{table*}

\clearpage
\listoftodos

\end{document}